%% file: nat_phys_resubmission/main_resubmission.tex
\documentclass[reprint,amsmath,amssymb,aps,prx,nofootinbib,superscriptaddress]{revtex4-2}
\usepackage{graphicx,siunitx,hyperref,orcidlink}
\renewcommand{\url}[1]{\href{#1}{[link]}}
\newcommand{\WIS}{Department of Condensed Matter Physics, Weizmann Institute of Science, Rehovot, Israel}
\newcommand{\IITJ}{Department of Physics, Indian Institute of Technology Jammu, Jammu 181221, India}
\newcommand{\UGA}{Univ. Grenoble Alpes, CEA, Grenoble INP, IRIG, PHELIQS, 38000 Grenoble, France}
\newcommand{\LAN}{Department of Physics, Lancaster University, Lancaster LA1 4YB, United Kingdom}
\newcommand{\ket}[1]{\left| #1 \right\rangle}
\newcommand{\bra}[1]{\left\langle #1 \right|}
\newcommand{\ketbra}[2]{\left| #1 \right\rangle \! \left\langle #2 \right|}
\newcommand{\Tint}{T_\text{int}}
\input{lambda_macros}

\newcommand{\fullref}[2]{\hyperref[#1]{\ref*{#1}#2}}

\begin{document}

\title{Gradually opening Schrödinger's box reveals a cascade of sharp dynamical transitions}

\author{Barkay Guttel\normalfont\,\orcidlink{0000-0003-0121-218X}}
\thanks{These authors contributed equally to this work.}
\affiliation{\WIS}

\author{Danielle Gov}
\thanks{These authors contributed equally to this work.}
\affiliation{\WIS}

\author{Noam Netzer}
\affiliation{\WIS}

\author{Uri Goldblatt}
\affiliation{\WIS}

\author{Sergey Hazanov}
\affiliation{\WIS}

\author{Lalit M. Joshi}
\affiliation{\WIS}

\author{Alessandro Romito\normalfont\, \orcidlink{0000-0003-3082-6279}}
\affiliation{\LAN}

\author{Yuval Gefen}
\affiliation{\WIS}

\author{Parveen Kumar\normalfont\,\orcidlink{0000-0003-3132-203X}}
\affiliation{\IITJ}

\author{Kyrylo Snizhko\normalfont\,\orcidlink{0000-0002-7236-6779}}
\affiliation{\UGA}

\author{Fabien Lafont}
\affiliation{\WIS}

\author{Serge Rosenblum\normalfont\,\orcidlink{0000-0002-1984-3206}}
\email{serge.rosenblum@weizmann.ac.il}
\affiliation{\WIS}

\date{\today}

\begin{abstract}
Quantum mechanics predicts that unobserved systems may exist in a superposition of states,
yet measurement produces definite outcomes—a tension at the heart of the quantum-to-classical boundary.
How the transformation between these opposing regimes unfolds as observation strength increases has remained experimentally elusive.
Here, by continuously tuning the measurement
strength on a superconducting qubit, we reveal that measurement-dominated dynamics emerge not
gradually but through multiple sharp transitions marked by the halting of coherent oscillations, state freezing near an attracting state, and the onset of quantum Zeno dynamics.
Decoherence, rather than washing out this structure, reshapes it---reordering the transitions and decoupling signatures that coincide in idealised models.
These results establish that the passage from quantum dynamics to measurement-dominated behaviour unfolds through a sequence of distinct transitions governed by the interplay between observation and environment.
\end{abstract}

\maketitle

Measurement shapes the boundary between quantum and classical physics. Unobserved quantum systems evolve continuously under the Schrödinger equation, but measurement forces them into definite states.
The tension between these two forms of evolution has driven decades of research on monitored quantum systems.
Experiments have resolved measurement backaction and individual quantum trajectories~\cite{sauter_observation_1986, bergquist_observation_1986,nagourney_shelved_1986,vijay_observation_2011,hacohen-gourgy_quantum_2016, hatridge_quantum_2013,murch_observing_2013,weber_mapping_2014, ficheux_dynamics_2018, gleyzes_quantum_2007, campagne-ibarcq_observing_2016}, explored non-Hermitian evolution in postselected dynamics~\cite{naghiloo_quantum_2019,wu_observation_2019, ding_experimental_2021, dogra_quantum_2021, chen_quantum_2021}, and observed measurement-induced phase transitions in many-body systems~\cite{noel_measurement-induced_2022,koh_measurement-induced_2023, hoke_measurement-induced_2023}.

\begin{figure}[b!]
\vspace{-20pt}
    \centering
     \includegraphics[width=.94\linewidth]{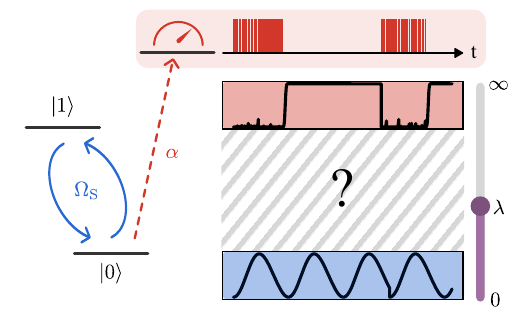}
    \setlength{\abovecaptionskip}{-10pt}
    \caption{
    \textbf{Monitored qubit with tunable measurement strength.}
    The qubit is resonantly driven at a rate $\Omega_\mathrm{S}$ (blue arrows) while being continuously monitored by a detector. The detector produces clicks (red bars) at a rate $\alpha$ whenever the qubit is in the ground state $\ket{0}$.
    The competition between coherent driving and measurement backaction is controlled by the dimensionless measurement strength $\lambda \equiv \alpha / (2\Omega_\mathrm{S})$. Increasing $\lambda$ takes the dynamics from Rabi oscillations to jump-like trajectories, illustrated by the simulated excited-state population traces (blue and red shaded, respectively).
    }
    \label{fig:intro_figure}
\end{figure}

A two-level system with tunable monitoring strength offers a precisely controllable arena for exploring this physics.  Consider a qubit with ground state $\ket{0}$ and excited state $\ket{1}$, driven resonantly at a Rabi rate $\Omega_\mathrm{S}$ and monitored by a detector that clicks at a rate $\alpha$ whenever the system is in $\ket{0}$ (Fig.~\ref{fig:intro_figure}).
The competition between coherent driving and
measurement backaction is then governed by a single dimensionless parameter, the
measurement strength $\lambda \equiv 
\alpha / (2 \Omega_\mathrm{S})$.
Quantum trajectory theory~\cite{carmichael_quantum_1993, molmer_monte_1993, wiseman_quantum_1996,dalibard_wave-function_1992, korotkov_continuous_1999, korotkov_selective_2001} resolves this competition at the
level of individual runs. Stochastic detector clicks project the qubit onto $\ket{0}$, while between clicks, the wavefunction evolves deterministically under a non-Hermitian Hamiltonian (Supplementary Section~\ref{App:ideal_model})
\begin{equation}
\label{Eq:non-hermitian hamiltonian}
    \hat{H}_{\text{eff}} / \hbar =\frac{\Omega_{\mathrm{S}}}{2}
    \left( \hat{\sigma}_\mathrm{y} -2i \lambda \ketbra{0}{0} \right).
\end{equation}
This no-click evolution is conditioned on a particular measurement record and is therefore a form of postselection, analogous to survival postselection in lossy systems~\cite{naghiloo_quantum_2019} and related to, but distinct from, weak-value protocols~\cite{dressel_colloquium_2014}.
The limiting regimes are well understood. For weak measurements ($\lambda\ll1$), the Rabi term dominates, and the qubit oscillates between $\ket{0}$ and $\ket{1}$, its rotation occasionally punctuated by clicks (Fig.~\ref{fig:intro_figure}).
For strong measurements ($\lambda\gg 1$), the anti-Hermitian term takes over: each interval without a click carries information about the state~\cite{renninger_messungen_1960}, suppressing the amplitude
of $\ket{0}$ relative to $\ket{1}$. Sustained silence therefore
steers the conditioned state towards the excited state---a
``continuous quantum jump'' that has been resolved
experimentally~\cite{minev_catch_2019}.
In the same limit, stronger measurement paradoxically slows the qubit's relaxation towards its steady state --- the quantum Zeno effect~\cite{itano_quantum_1990, fischer_observation_2001, kakuyanagi_observation_2015, xiao_nmr_2006, palacios-laloy_experimental_2010, slichter_quantum_2016, harrington_quantum_2017}.

\begin{figure*}[ht!]
    \centering
    \includegraphics[width=0.9\textwidth]{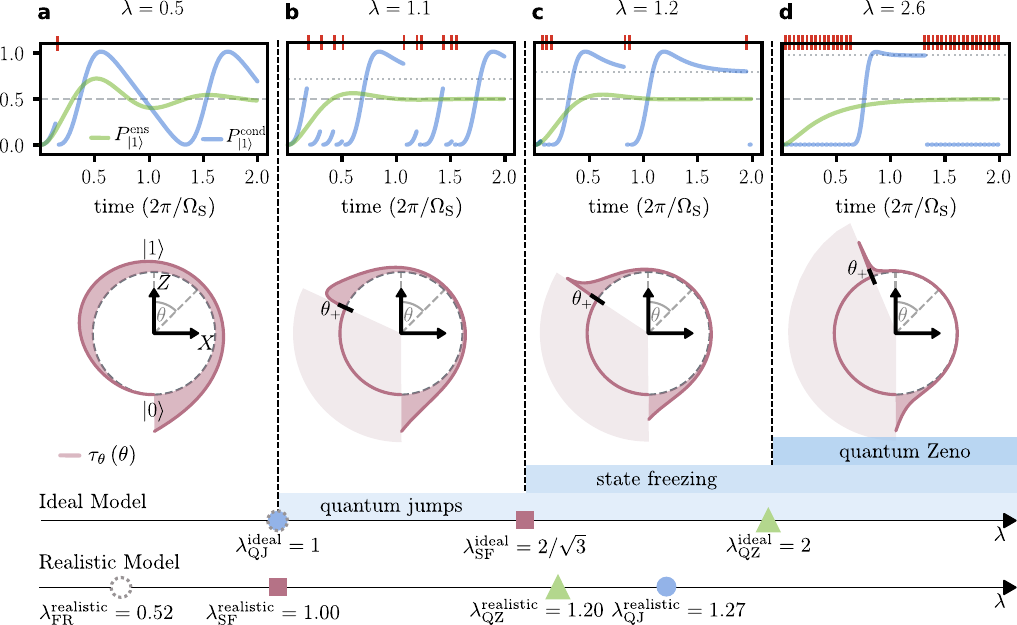}
    \setlength{\belowcaptionskip}{-15pt}
    \setlength{\abovecaptionskip}{5pt}
    \caption{
\textbf{Dynamical transitions of a monitored qubit, and their reordering by experimental imperfections.}
Simulated dynamics of the ideal monitored qubit at four measurement strengths $\lambda \equiv \alpha / (2 \Omega_\mathrm{S})$, representing the four dynamical regimes. The three transitions occur at $\critlambda{1}=1$ (onset of continuous quantum jumps), $\critlambda{2}=2/\sqrt{3}$ (state freezing) and $\critlambda{3}=2$ (quantum Zeno effect). 
\emph{Top:} A representative binary click record (red ticks), the excited-state probability $P_{\ket{1}}^{\mathrm{cond}}$ (blue) of the corresponding trajectory, which is reset to $\ket{0}$ at each click, and the ensemble average $P_{\ket{1}}^{\mathrm{ens}}$ (green). The qubit is initialised in $\ket{0}$ and reaches a steady-state probability of $1/2$ (grey dashed line).
\emph{Middle:} In the ideal model, the polar angle $\theta$ uniquely parametrises the trajectory. The angular dwell time density $\tau_\theta(\theta)$ is shown in rose as a radial height above the unit circle (a.u.). Shading marks the forbidden region, and $\theta_+$ is the attracting fixed-point angle.
\textbf{a,} Below $\critlambda{1}$, the no-click evolution is oscillatory and visits every angle.
\textbf{b,} Above $\critlambda{1}$, the no-click oscillations cease and the evolution becomes a continuous quantum jump towards $\theta_+$ (grey dotted line), which opens a forbidden region on the Bloch sphere.
The jump is typically interrupted by a click before reaching $\theta_+$.
\textbf{c,} Above $\critlambda{2}$, trajectories can reach $\theta_+$ before a click and remain there long enough for $\tau_\theta$ to diverge rather than vanish.
\textbf{d,} Above $\critlambda{3}$, the ensemble dynamics become overdamped and further increasing $\lambda$ slows relaxation towards the steady state.
\emph{Bottom:} the predicted ordering of the transitions in the ideal model (upper axis) and in the realistic model corresponding to our experiment, which includes decoherence and detector imperfections (lower axis; Supplementary Section~\ref{App:realistic_model}).
Decoherence and detector imperfections decouple the opening of the forbidden region $\reallambda{4}$ (dotted open marker) from the cessation of oscillations at $\reallambda{1}$, and reorder the transitions such that state freezing and the onset of quantum Zeno dynamics both precede the quantum-jump transition ($\reallambda{2}<\reallambda{3}<\reallambda{1}$).
Unlike the other thresholds, $\reallambda{4}$ is a geometric feature of the trajectory that depends on the initial state, rather than a spectral transition.
}
    \label{fig:schematic_and_transitions}
\end{figure*}

A central question is the nature of the crossover between these two regimes---whether it is abrupt or gradual, featureless or rich. Theoretical work predicted that features of coherent dynamics are not lost gradually, but in sharp transitions, each unveiling qualitatively distinct dynamics and statistics of quantum trajectories.
Such transitions have been identified in null-result-conditioned dynamics~\cite{ruskov_crossover_2007}, quantum-jump statistics~\cite{ren_braid_2013,li_quantum_2014,pavlov_topological_2025}, and repetitive ancilla-based measurements~\cite{popperl_localization_2024}.
In particular, ref.~\cite{snizhko_quantum_2020} predicted that, for an ideal qubit free of decoherence, the coherent and Zeno regimes are connected by a cascade of three transitions (Fig.~\ref{fig:schematic_and_transitions}). At an exceptional point~\cite{ashida_non-hermitian_2020,minganti_quantum_2019} of the conditional no-click Hamiltonian $\hat{H}_\text{eff}$, occurring at $\critlambda{1} = 1$, coherent oscillations halt. Instead, trajectories undergo continuous quantum jumps towards a stable state~\cite{minev_catch_2019,katz_coherent_2006}, and a forbidden region emerges on the Bloch sphere.
Beyond $\critlambda{2} = 2/\sqrt3$, the dwell time near that stable state diverges and trajectories freeze in its vicinity.
Finally, at $\critlambda{3} = 2$, an exceptional point of the ensemble-averaged Liouvillian, the qubit enters the quantum Zeno regime~\cite{peres_zeno_1980,misra_zenos_1976,gambetta_quantum_2008,greenfield_unified_2025,hacohen-gourgy_incoherent_2018}.

\noindent Here, we resolve this cascade experimentally. We observe the predicted transitions as we tune the measurement strength, but not in the predicted order: decoherence and detector imperfections decouple the opening of the forbidden region from the cessation of oscillations and reorder the cascade such that state freezing and the onset of quantum Zeno dynamics both precede the quantum-jump transition~(Fig.~\ref{fig:schematic_and_transitions}, bottom). Decoherence and detector imperfections therefore do not blur the cascade but fundamentally reorganise it.

\begin{figure*}
    \centering
    \includegraphics[width=\linewidth]{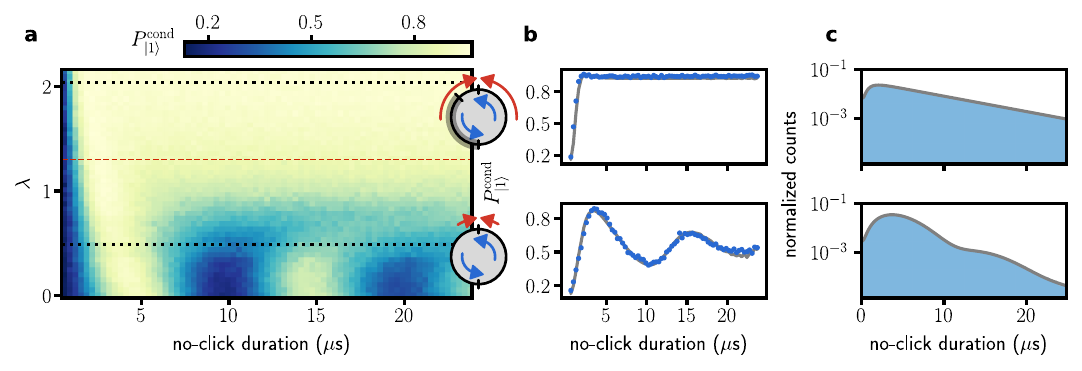}
    \vspace{-17pt}
    \caption{
    \textbf{Observation of the transition to the quantum-jump regime.} 
     Below the quantum-jump transition (bottom Bloch sphere), coherent driving (blue arrows) dominates measurement-induced backaction (red arrows) across the entire Bloch sphere, resulting in oscillations. Above it (top Bloch sphere), measurement backaction precisely balances the coherent drive at a particular polar angle, causing oscillations to halt and giving rise to quantum-jump dynamics.
     \textbf{a,} Measured excited-state population $P_{\ket{1}}^{\mathrm{cond}}$ as a function of the no-click sequence duration and measurement strength $\lambda$, revealing the shift from oscillatory to jump-like behaviour.
     The dashed red line marks $\obslambdafull{1}$, extracted from the no-click survival probability (Supplementary Fig.~\ref{fig:crit_lambda_1_with_poles}). Just below this value, the oscillations are too strongly damped to be visible.
     Each data point is the average of \num{1500} experimental runs.
     \textbf{b,} One-dimensional cuts below (bottom panel, $\lambda=0.5$) and above (top panel, $\lambda=2.1$) the quantum-jump transition at $\lambda$ values indicated by the dotted lines in~\textbf{a}.
     Grey lines show simulations of the no-click dynamics using independently measured decoherence rates (Supplementary Section~\ref{App:numerical_sim}).
     Shot-noise error bars are smaller than the marker size.
     \textbf{c,} Histograms of no-click durations at the same measurement strengths as in \textbf{b}. Grey solid lines show fits using a sum of three complex exponentials  (Methods).
    The qubit is initialised in \(\ket{0}\) for \textbf{a,b}, whereas in \textbf{c} it is initialised in \(\ket{1}\) to suppress the effect of detection errors in long no-click intervals.
    The single-measurement integration time is $\Tint=\qty{200}{\nano\second}$ in~\textbf{c},  compared with $\Tint=\qty{320}{\nano\second}$ for the real-time experiments in~\textbf{a,b}. 
    Throughout, the coherent drive strength is set at $\Omega_\mathrm{S}/2\pi=\qty{100}{\kilo\Hz}$. 
    }
    \vspace{-10pt}
    \label{fig:first_transition}
\end{figure*}

We realise the monitored qubit using a superconducting circuit consisting of a system qubit coupled to an ancillary detector qubit, forming a V‑type three‑level system similar to ref.~\cite{minev_catch_2019}.
We control the measurement strength $\lambda$ by varying the amplitude of a drive resonant with the detector qubit transition (see Methods).
Our analysis combines binary detector click records with quantum state tomography, performed both unconditionally and conditioned on the detector record. The latter resolves dynamical features that ensemble averages obscure. Together, these methods yield a map of the dynamical phases of a monitored qubit.

\begin{figure*}[t!]
\centering
\includegraphics[width=\linewidth]{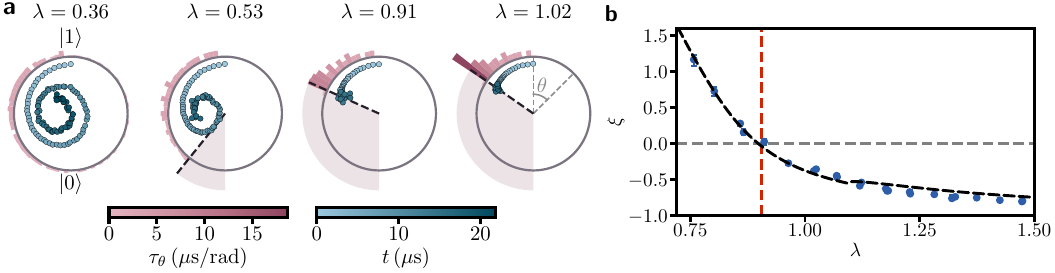}
\setlength\abovecaptionskip{0pt}
\setlength\belowcaptionskip{-10pt}
    \caption{
\textbf{Emergence of the forbidden region and transition into the state-freezing regime.}
\textbf{a,} Conditional Bloch-vector trajectories (circular markers) reconstructed by tomography after initialisation in $\ket{1}$, at four representative measurement strengths $\lambda$. Trajectories at additional values of $\lambda$ are shown in Supplementary Fig.~\ref{fig:second_transition_bloch_grid}. Colour denotes evolution time. Readouts are shown while the no-click probability $P^{(0)}(t)\geq0.01$.
Bars outside the unit circle show the binned dwell time per unit angle $\tau_\theta(\theta)$ (Methods), with an angular bin width of $2\pi/80$.
The colour bar indicates the angular dwell-time density.
At $\lambda=0.36$, the trajectory completes full Rabi oscillations and visits every polar angle. As $\lambda$ increases, it spirals into an attracting fixed point without completing a full winding, opening a forbidden region (shaded sectors) well before oscillations cease.
Above the state-freezing transition, at $\lambda=1.02$, the angular dwell time is strongly concentrated near the attracting fixed point, characteristic of the state-freezing regime.
\textbf{b,}~Critical exponent $\xi=\gamma_0/\gamma_\mathrm{a}-1$ versus $\lambda$, where $\gamma_\mathrm{a}$ is the late-time approach rate and $\gamma_0$ is the asymptotic decay rate of the no-click survival probability. Shot-noise error bars are $\qty{68}{\percent}$ bootstrap intervals. The vertical red line marks the observed state-freezing transition ($\xi=0$) at $\obslambdafull{2}$, obtained from a local quadratic fit 
  (Supplementary Section~\ref{App:extracting_critical_lambdas}). The dashed black curve shows $\xi(\lambda)$ calculated from the realistic model, shifted horizontally to compare its shape with the data. Throughout, $\Omega_\mathrm{S}/2\pi=\qty{100}{\kilo\Hz}$.}
    \label{fig:second_transition}
\end{figure*}

\vspace{-10pt}
\subsection*{Transition into the quantum‑jump regime}
\label{sec:first_transition}
\vspace{-5pt}
The transition into the quantum-jump regime arises from competition between two coherent processes acting during no-click evolution (equation~\eqref{Eq:non-hermitian hamiltonian}): the Rabi drive attempts to rotate the qubit around the Bloch sphere, while the no-click measurement backaction pulls it towards the excited state $\ket{1}$.
Below a critical measurement strength $\lambda < \critlambda{1}$, Rabi driving dominates at all polar angles \(\theta\) on the Bloch sphere, allowing the qubit to complete full periodic rotations. As $\lambda$ approaches $\critlambda{1}$, the oscillations become increasingly asymmetric, reflecting the growing impact of measurement backaction (Fig.~\fullref{fig:schematic_and_transitions}{a}).
In the ideal model, the critical point $\critlambda{1}=1$ marks the threshold at which measurement backaction first becomes strong enough to halt the qubit's rotation. For any $\lambda \geq \critlambda{1}$, backaction precisely balances coherent driving at an angle $\theta_{+}(\lambda)=2\arctan \! \left(\sqrt{\lambda^2-1}-\lambda\right)$, creating an attracting fixed point\footnote{
A second fixed point appears at $\theta_{-}\equiv -\pi -\theta_{+}$~\cite{snizhko_quantum_2020}. However, this fixed point is unstable, so the qubit reaches it only if specifically initialised in that state.
}~\cite{ruskov_crossover_2007}.
Trajectories now execute continuous, deterministic quantum jumps from $\ket{0}$ towards the eigenstate of $\hat{H}_\text{eff}$, $\ket{\psi_+}\equiv\sin \! \left(\frac{\theta_{+}}{2}\right)\ket{0}+\cos \! \left(\frac{\theta_{+}}{2}\right)\ket{1}$, until a stochastic detector click resets the qubit to $\ket{0}$ (Fig.~\fullref{fig:schematic_and_transitions}{b}). The fixed point at $\ket{\psi_+}$ creates a forbidden region $(-\pi,\theta_{+}(\lambda)]$ on the Bloch sphere, which eventually encompasses the entire left hemisphere as measurement strength increases~\cite{snizhko_quantum_2020} (see the Bloch spheres in Figs.~\fullref{fig:schematic_and_transitions}{b--d}).

This transition occurs at an exceptional point of the non-Hermitian no-click Hamiltonian $\hat{H}_\mathrm{eff}$, at which its eigenvalues and eigenvectors coalesce. Since $\hat{H}_\mathrm{eff}$ is parity-time ($\mathcal{PT}$) symmetric, the exceptional point separates unbroken and broken $\mathcal{PT}$-symmetric phases~\cite{bender_making_2007,ashida_non-hermitian_2020}. Beyond this point, the eigenvalue difference becomes purely imaginary, causing the oscillation frequency to vanish.
This exceptional point is a property of the conditional no-click evolution, rather than of the stochastic click-and-reset dynamics. This type of conditioning parallels the dynamics of lossy non-Hermitian systems, which are known to exhibit comparable exceptional-point transitions~\cite{naghiloo_quantum_2019, wu_observation_2019, ding_experimental_2021, dogra_quantum_2021, wang_observation_2021, hu_full-parameter-space_2025}.

To observe this transition experimentally, we reconstruct the qubit's evolution conditioned on the absence of detector clicks. We measure the qubit's excited-state population after detecting in real time a no-click sequence of the desired duration. Repeating this procedure across different no-click durations and measurement strengths reveals a shift from oscillatory to quantum-jump dynamics (Figs.~\fullref{fig:first_transition}{a,b}).
The oscillations become increasingly distorted as $\lambda$ approaches the critical value. The observed no-click dynamics are in excellent qualitative agreement with numerical simulations that incorporate decoherence (Supplementary Section~\ref{App:numerical_sim}).

The distribution of no-click durations confirms this transition independently. Initialising the system in $\ket{1}$, we record the time to the first detector click.
Since the instantaneous click rate varies as $r(\theta)\equiv\alpha\sin^2 \! \left(\frac{\theta(t)}{2}\right)$, Rabi oscillations imprint a periodic modulation on this distribution. The data (Fig.~\fullref{fig:first_transition}{c}) confirm this expectation: oscillatory patterns at weak measurements give way to monotonic decay as $\lambda$ increases.
From the vanishing of oscillations we locate the transition at $\obslambdafull{1}$ (Methods and Supplementary Section~\ref{App:extracting_critical_lambdas}),
consistent with $\reallambda{1}=\reallambdaval{1}$, the prediction of a realistic model that includes decoherence and detector imperfections (Supplementary Section~\ref{App:transition_locations_The_First_Transition}).

Full state tomography conditioned on no clicks resolves the entire Bloch sphere trajectory, rather than the excited-state population alone (Fig.~\fullref{fig:second_transition}{a}). At weak measurement strengths (\(\lambda = 0.36\)), the reconstructed trajectory completes full Rabi oscillations, visiting all polar angles. Intriguingly, as $\lambda$ increases, a forbidden angular region emerges well before oscillations cease. This decoupling of two signatures that coincide in the ideal model is a consequence of decoherence: the conditioned state becomes mixed, so the trajectory spirals through the interior of the Bloch sphere towards an attracting fixed point.
At small $\lambda$, the spiral still encircles the centre, so all polar angles remain visited. A forbidden region opens once the spiral converges without completing a full winding.
Unlike the cessation of oscillations, this threshold does not result from an underlying spectral degeneracy, but instead marks a geometric change in the trajectory for a particular initial condition. The observed opening of a forbidden region occurs at $\obslambdafull{4}$, slightly below the realistic model prediction of $\reallambda{4} = \reallambdaval{4}$ for a trajectory initialised in $\ket{1}$.
We discuss the fundamental role of decoherence in enabling the splitting of the two features --- cessation of oscillations and opening of the forbidden region --- in more detail in Supplementary Section~\ref{App:first_transition_splitting}.

\vspace{-10pt}
\subsection*{Transition into the state-freezing regime}
\label{sec:second_transition}
\vspace{-5pt}

A hallmark of strongly measured quantum systems is state freezing, in which the qubit does not merely approach the attracting fixed point but remains trapped in its vicinity for extended periods.
Two competing processes determine whether this happens. As a no-click trajectory approaches the fixed point, its motion slows, increasing the time spent nearby. At the same time, detector clicks reset the qubit to the ground state. State freezing sets in when the slowing motion causes trajectories to accumulate near the fixed point faster than clicks remove them.

We quantify this competition through the dwell-time density along the no-click trajectory. Because the experimentally observed trajectories spiral through the Bloch sphere, the polar angle \(\theta\) does not uniquely specify a position along the path. We therefore parametrise the trajectory by its arc length \(l\), defining \(\tau_l(l)\,dl\) as the mean time spent in an interval \(dl\). Near the attracting fixed point, at arc length \(l_\infty\), the dwell-time density scales as 

\begin{equation*}
{\tau_l(l)\propto(l_\infty-l)^\xi,\qquad
\xi=\frac{\gamma_0}{\gamma_{\mathrm a}}-1,}
\end{equation*}
where \(\gamma_{\mathrm a}\) is the rate of approach to the fixed point and \(\gamma_0\) is the decay rate of the no-click survival probability (Supplementary Section~\ref{App:extracting_second_transition_lambda}). For $\xi>0$, trajectories are reset by clicks faster than they approach the fixed point, and the dwell-time density vanishes there. For $\xi<0$, the approach dominates and the dwell-time density diverges. The state-freezing transition occurs when the two rates become equal, $\gamma_{\mathrm a}=\gamma_0$. It is therefore governed by a balance of dynamical rates rather than an exceptional point. In the ideal model, this transition occurs at \(\critlambda{2}=2/\sqrt{3}\approx1.15\)~\cite{snizhko_quantum_2020}.

The angular dwell-time distributions in Fig.~\fullref{fig:second_transition}{a} illustrate this accumulation, with a pronounced concentration near the attracting fixed point at $\lambda = 1.02$.
To locate the transition quantitatively, we extract the approach rate $\gamma_\mathrm{a}$ from the late-time conditional tomography and the survival decay rate $\gamma_0$ from the no-click probability. Their ratio yields \(\xi(\lambda)\), whose zero crossing identifies the state-freezing transition at \(\obslambdafull{2}\), compared with the realistic-model prediction \(\reallambda{2}=\reallambdaval{2}\) (Fig.~\fullref{fig:second_transition}{b}; Methods and Supplementary Section~\ref{App:transition_locations_The_Second_Transition}). 

Strikingly, the state-freezing transition is observed below the quantum-jump transition, reversing the ordering of the ideal model. This inversion again reflects the influence of decoherence.
In the ideal model, an attracting fixed point appears only once oscillations cease, so state freezing can occur only after the quantum-jump transition. With decoherence, however, the no-click trajectory spirals towards an attracting fixed point inside the Bloch sphere while remaining oscillatory (Fig.~\fullref{fig:second_transition}{a}), removing this constraint. The ideal ordering is therefore no longer enforced. The physical origin of the inversion can be understood from how dephasing affects the two transitions differently.
Phase flips can enable trajectories to complete a cycle around the Bloch sphere without
a click being recorded, thus sustaining oscillations at higher \(\lambda\) and shifting the quantum-jump transition upwards. At the same time, dephasing produces a contraction towards the mixed-state fixed point, increasing the rate of approach relative to the late-time click rate and causing state freezing to occur at smaller $\lambda$. Perturbative and numerical analysis confirms these opposite shifts and the resulting crossing of the transition locations (Fig.~\ref{fig:schematic_and_transitions} and Supplementary Section~\ref{App:transition_locations}).

\begin{figure}[h!]
    \centering
    \includegraphics[width=0.95\linewidth]{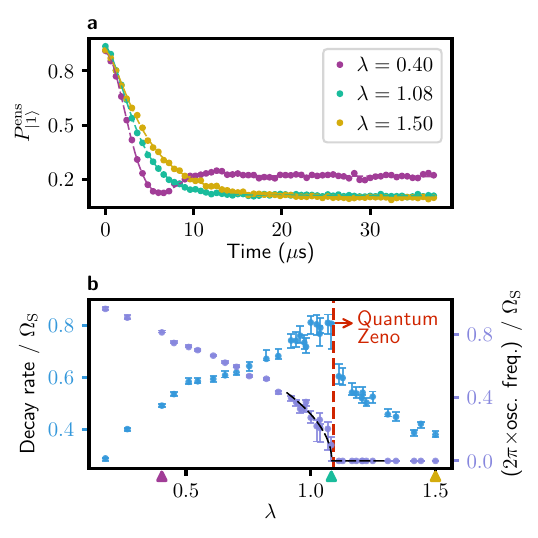}
    \caption{
    \textbf{Relaxation to the steady state across the quantum Zeno transition.}
    At this transition, oscillatory decay gives way to overdamped relaxation; beyond it, stronger measurement slows relaxation.
    \textbf{a,} Ensemble-averaged excited-state probability $P_{\ket{1}}^\mathrm{ens}$ following initialisation in $\ket{1}$, for values of $\lambda$ chosen to illustrate behaviour below, near, and above the transition.
    Each data point is the average of \num{3000} experimental runs. Error bars are smaller than the marker size.
    Dashed lines show fits using a sum of three complex exponentials.
    The steady-state value of $P_{\ket{1}}^\mathrm{ens}$ decreases with $\lambda$ because a growing fraction of the population resides in the detector's excited state.
    \textbf{b,} Extracted relaxation rates (blue) and oscillation frequencies (purple) versus \(\lambda\). Error bars indicate $\qty{68}{\%}$ bootstrap confidence intervals.
    Coloured triangles mark the measurement strengths shown in \textbf{a}. The steady-state value in \textbf{a} only affects the constant term of each fit and therefore does not bias these rates and frequencies.
   The dashed black line shows the oscillation frequency obtained from a fit to the predicted behaviour near the transition (Methods).
     The transition is observed at $\obslambdafull{3}$ (dashed red line), below the ideal value $\critlambda{3}=2$ owing to the finite waiting time in the detector's excited state.
    Throughout, the coherent drive is \(\Omega_\mathrm{S} / 2\pi = \qty{100}{\kilo\Hz}\).
    }
    \vspace{-10pt}
    \label{fig:third_transition}
\end{figure}

\vspace{-10pt}
\subsection*{Transition into the quantum Zeno regime}
\label{sec:third_transition}
\vspace{-5pt}

Unlike the quantum-jump and state-freezing transitions, the quantum Zeno transition manifests in the unconditional ensemble dynamics rather than the conditional evolution~\cite{li_quantum_2014, snizhko_quantum_2020, greenfield_unified_2025}. Beyond this point, increasing the measurement strength slows rather than accelerates the relaxation towards the steady state, as observed in related quantum systems~\cite{fischer_observation_2001, kakuyanagi_observation_2015, xiao_nmr_2006, palacios-laloy_experimental_2010} (Figs.~\fullref{fig:schematic_and_transitions}{c,d}).
In addition, measurement-induced dephasing critically damps the ensemble dynamics at the transition point, turning decaying oscillations into an overdamped decay.
These changes stem from an exceptional point of the Liouvillian superoperator $\mathcal{L}$ 
governing the evolution of the ensemble-averaged density matrix $\hat{\rho}(t)$~\cite{minganti_quantum_2019, dubey_quantum_2023},
\begin{equation}
    \label{Eq:full_liouvillian}
    \dot{\hat{\rho}} = \mathcal{L}\hat{\rho} = -i \left[\frac{\Omega_\mathrm{S}}{2}\hat{\sigma}_\mathrm{y}, \hat{\rho}(t)\right]
    + \alpha \mathcal{D}\left[\ketbra{0}{0}\right] \hat{\rho}(t),
\end{equation}
with the dissipator \(\mathcal{D}[\hat{L}]\hat{\rho} = \hat{L} \hat{\rho} \hat{L}^{\dagger} - \frac{1}{2} \left\{ \hat{L}^{\dagger}\hat{L}, \hat{\rho} \right\}\) capturing both the stochastic clicks generated by \(\hat{L} = \ketbra{0}{0}\) and the no-click backaction generated by $\hat{L}^{\dagger}\hat{L}$.

\begin{figure}
    \centering 
    \includegraphics[width=0.85\linewidth]{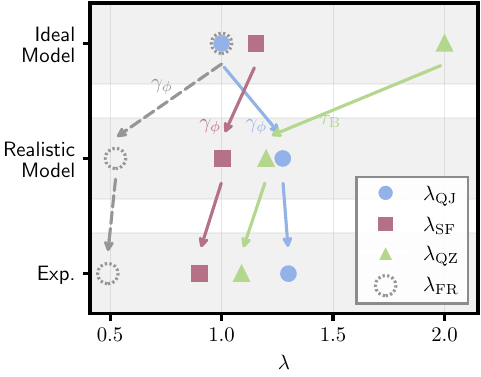}
    \caption{\textbf{Summary of transition locations.}
    Comparison of the three transitions $\generallambda{1}$, $\generallambda{2}$ and $\generallambda{3}$ (filled markers) and the forbidden-region threshold $\lambda_{\rm FR}$ (dotted open marker) across three levels of description: the ideal model ($\critlambda{1} = \critlambda{4} = 1$, $\critlambda{2} = 2/\sqrt{3}$, $\critlambda{3} = 2$), the realistic model incorporating decoherence and detector imperfections ($\reallambda{1} = \reallambdaval{1}$, $\reallambda{2} = \reallambdaval{2}$, $\reallambda{3} = \reallambdaval{3}$,
    $\reallambda{4} = \reallambdaval{4}$; Supplementary Section~\ref{App:transition_locations}), and experimental observations ($\obslambdafull{1}$, $\obslambdafull{2}$, $\obslambdafull{3}$,
    $\obslambdafull{4}$; Supplementary Section~\ref{App:extracting_critical_lambdas}).
    Notably, decoherence and detector imperfections reorder the transitions, a feature captured by the realistic model ($\reallambda{2} < \reallambda{3} < \reallambda{1}$).
    Each arrow linking the ideal to the realistic model is labelled with the parameter that dominates that shift: pure dephasing $\gamma_\phi$ for $\generallambda{1}$, $\generallambda{2}$ and $\generallambda{4}$, and the finite waiting time $\tau_\mathrm{B}$ for $\generallambda{3}$.
    The statistical uncertainties on the observed values are smaller than the marker size.
    The observed
$\obslambda{2}$ and $\obslambda{3}$ lie about \qty{10}{\%} below the realistic-model
values, well outside these uncertainties. This discrepancy may reflect calibration and
systematic effects not captured by the model.}

    \label{fig:summary}
\end{figure}

To probe this transition, we perform state tomography after variable evolution times without conditioning on the click record (Fig.~\fullref{fig:third_transition}{a}).
As $\lambda$ increases, the relaxation rate initially increases while the oscillation frequency decreases (Fig.~\fullref{fig:third_transition}{b}).
However, at the observed transition point $\obslambdafull{3}$ (Methods), the oscillations vanish and the relaxation becomes overdamped.
At still larger \(\lambda\), the relaxation rate decreases, displaying the characteristic Zeno slowdown.
The observed transition lies substantially below the ideal prediction \(\critlambda{3}=2\), and even below the observed quantum-jump transition \(\obslambda{1}\). The realistic model predicts $\reallambda{3}=\reallambdaval{3}$, with the downward shift arising predominantly from the finite waiting time \(\tau_\mathrm{B}\) in the detector's excited state following a click (Supplementary Section~\ref{App:transition_locations_The_Third_Transition}). This sensitivity to \(\tau_\mathrm{B}\) reflects the fact that, unlike the other two transitions, the quantum Zeno transition depends on the resetting dynamics following detector clicks.
Varying \(\tau_\mathrm{B}\) interpolates between the ideal monitored limit, in which the system returns instantaneously to \(\ket{0}\) after a click \((\tau_\mathrm{B}\to0,\ \generallambda{3}=2)\), and the postselected limit, in which the system never returns from the detector's excited state \((\tau_\mathrm{B}\to\infty,\ \generallambda{3}\to\generallambda{1})\).
Dephasing further shifts the quantum Zeno and quantum-jump transitions in opposite directions, lowering the former and raising the latter, and detector imperfections raise the quantum-jump transition further. Together with the finite detector waiting time, this explains their inverted ordering in the realistic model (Supplementary Section~\ref{App:transition_locations}). This ordering agrees with experiment, although the measured state-freezing and quantum Zeno thresholds are approximately $\qty{10}{\%}$ lower than predicted (Fig.~\ref{fig:summary}, Supplementary Section~\ref{App:model_vs_experiment}).

\subsection*{Conclusions}

The route from drive-dominated coherent evolution to measurement-dominated behaviour unfolds not as a gradual crossover, but as a cascade of dynamical transitions, each revealing a distinct hallmark of monitored quantum systems.
The three transitions probe complementary aspects of the monitored dynamics: the quantum-jump transition is a property of the conditional no-click evolution, the state-freezing transition additionally depends on the no-click survival probability of those trajectories, and the quantum Zeno transition appears in the unconditional ensemble dynamics.
Strikingly, decoherence fundamentally restructures the dynamical phase diagram (Fig.~\ref{fig:summary}). Dephasing lowers the state-freezing threshold while raising the quantum-jump threshold, whereas the detector's finite waiting time lowers the quantum Zeno threshold. As a result, both state freezing and quantum Zeno dynamics emerge before the onset of continuous quantum jumps, reversing their ordering in the ideal model (Supplementary Section~\ref{App:transition_locations}). 

The structure uncovered here points to a broader landscape. Beyond measurement strength, the temporal spacing of observations has been predicted to drive its own set of transitions—between ergodic, fractal, and localised phases~\cite{popperl_localization_2024}.
Additional probes, such as click counting fields~\cite{li_quantum_2014} or trajectory-resolved entropy production~\cite{harrington_characterizing_2019}, may uncover further structure, including topologically distinct phases~\cite{li_quantum_2014,pavlov_topological_2025}.
More broadly, the cascaded emergence of measurement-induced dynamics should extend beyond the single-qubit setting, with implications for measurement-enhanced entanglement generation~\cite{li_speeding_2023} and measurement-induced phase transitions in many-body systems~\cite{skinner_measurement-induced_2019,li_quantum_2018,szyniszewski_entanglement_2019,fisher_random_2023}.

\clearpage
\bibliographystyle{naturemag}
\bibliography{references_zotero}

\clearpage

\section*{Methods}
\label{sec: methods}

\par
\textbf{Experimental setup.}
Our device comprises two fixed‑frequency transmon qubits housed in a 3D superconducting readout cavity~\cite{minev_catch_2019}. The system or ``dark'' qubit is oriented perpendicularly to the field of the cavity's fundamental mode, causing a weak dispersive shift of $\chi_\mathrm{S}/2\pi= \qty{0.27}{\MHz}$.
By contrast, the detector or ``bright'' qubit is aligned parallel to the field and couples an order of magnitude more strongly to the cavity ($\chi_\mathrm{B}/2\pi = \qty{5.1}{\MHz}$).
Due to the nonlinear interaction between the two qubits~\cite{mundada_suppression_2019}, the doubly excited state $\ket{e_\mathrm{S},e_\mathrm{B}}$ is shifted by $\chi_\mathrm{SB} / 2\pi = \qty{36}{\MHz}$ and becomes inaccessible, leaving an effective V-shaped three-level system: the ground state $\ket{0}\equiv\ket{g_\mathrm{S},g_\mathrm{B}}$, the dark excited state $\ket{1}\equiv\ket{e_\mathrm{S} ,g_\mathrm{B}}$ (system qubit excitation), and the bright excited state $\ket{\mathrm{B}}\equiv\ket{g_\mathrm{S},e_\mathrm{B}}$ (detector qubit excitation).

By applying a readout tone at $\omega=\omega_\mathrm{R}-\chi_\mathrm{B}$, with $\omega_\mathrm{R}$ the readout cavity frequency, we monitor the population in $\ket{\mathrm{B}}$ at a rate $\Gamma\propto \bar{n}_{\mathrm{B}}$, with $\bar{n}_{\mathrm{B}}$ the average number of readout photons set by the readout tone amplitude $\Omega_\mathrm{R}$.
Due to the large asymmetry in the dispersive shifts, we achieve rapid, single-shot readout distinguishing $\ket{\mathrm{B}}$ from $\ket{0}$ and $\ket{1}$ with an assignment fidelity exceeding \qty{99}{\percent} within an integration time $T_{\text{int}} = \qty{200}{\nano\second}$.
This integration time is used throughout the experiment, except for the real-time conditional measurements in Fig.~\fullref{fig:first_transition}{a,b}, where hardware constraints require $T_{\text{int}}=\qty{320}{\nano\second}$.
A weak resonant drive of amplitude $\Omega_\mathrm{B} \ll \Gamma$ induces excursions from $\ket{0}$ to $\ket{\mathrm{B}}$ at rate $\alpha\propto \Omega_\mathrm{B}^2/\Gamma$~\cite{slichter_quantum_2016}, which are registered as detector clicks when resolved by the readout.
To tune $\lambda$, we adjust $\Omega_\mathrm{B}$ while keeping the system drive $\Omega_\mathrm{S}$ and the readout tone  $\Omega_\mathrm{R}$ fixed.
More details on the experimental setup can be found in Supplementary Sections~\ref{App:Experimental Setup and Chip Fabrication} and~\ref{App:Hamiltonian and Lindbladian parameters}.

\par
\textbf{Calibration of the measurement rate $\alpha$.}
The measurement rate $\alpha$ is defined as the inverse of the mean waiting time the qubit spends in $\ket{0}$ (with $\Omega_\mathrm{S}=0$) before being excited to the detector's excited state $\ket{\mathrm{B}}$, where it produces a detector click.
We determine $\alpha$ from a hidden Markov model (HMM) analysis of the binary measurement record. In this approach, we use the Baum–Welch algorithm~\cite{baum_maximization_1970} to infer the underlying transition rate $\alpha$ from $\ket{0}$ to $\ket{\mathrm{B}}$ using the binary measurement record. This independently calibrated rate determines the dimensionless measurement strength via the relation $\lambda=\alpha/ (2\Omega_{\mathrm{S}})$. In addition to the measurement rate, the HMM analysis also provides the waiting time $\tau_\mathrm{B}$ in $\ket{\mathrm{B}}$.
More details are provided in Supplementary Section~\ref{App:click_statistics}.

\par
\textbf{Quantum state tomography.}
Conditional quantum state tomography at the end of an experimental run (Figs.~\ref{fig:first_transition} and \ref{fig:second_transition}) is performed only when a specified click pattern has been detected.
After switching off all three drives, a first projective measurement verifies that the system is not in the detection state $\ket{\mathrm{B}}$. Conditioned on finding the system within the qubit manifold $\{\ket{0},\ket{1}\}$, a set of calibrated Pauli operations maps the $\mathrm{X}$, $\mathrm{Y}$, or $\mathrm{Z}$ polarisations of the system manifold onto the detection manifold $\{\ket{0},\ket{\mathrm{B}}\}$.
A second projective measurement is then performed, enabling reconstruction of the density matrix of the system.
In contrast, unconditional tomography (Fig.~\ref{fig:third_transition}) is performed after a preset duration, irrespective of the measurement record. Runs in which the system is found in $\ket{\mathrm{B}}$ are retained, so that $P_{\ket{1}}^\mathrm{ens}$ is the probability of finding the system in $\ket{1}$ among all runs, rather than being renormalised to the qubit manifold.

\par
\textbf{Constructing angular dwell time histograms.}
To reconstruct the angular dwell-time distribution $\tau_{\theta}(\theta)$, we combine
two measurements: the no-click survival probability $P^{(0)}(t)$, obtained from the
click record as the fraction of runs with no click up to time $t$, and the angle
$\theta(t)$ of the no-click trajectory, obtained from conditional tomography.
From $P^{(0)}(t)$ we derive the first-click density
$f_t(t)\equiv-dP^{(0)}(t)/dt$, evaluated as the fraction of first clicks in each
readout window divided by $T_\mathrm{int}$, and the instantaneous click rate
$r(t)=f_t(t)/P^{(0)}(t)$.
Using $\theta(t)$, these are converted to their angular counterparts,
$f_\theta(\theta)$ and $r(\theta)$.
The mean dwell time per unit angle in the interval $[\theta, \theta + \Delta\theta)$
is then
\[
\tau_{\theta}(\theta) = \frac{f_\theta(\theta)}{r(\theta)} \,,
\]
as detailed in Supplementary Section~\ref{App:ideal_model}.

A narrow peak in the dwell time can fall on a bin edge and be split between two
adjacent bins, which reduces its apparent height by up to a factor of two. To avoid this binning artefact, we apply the same rule at every $\lambda$ in
Figs.~\fullref{fig:second_transition}{a} and~\ref{fig:second_transition_bloch_grid}:
we compute two histograms from the same data, one on a grid starting at $-\pi$
and one shifted by half a bin width, and display the one with the larger
single-bin dwell time within the window $\left(-\frac{\pi}{2},0\right)$, where
the trajectory accumulates near its fixed point.
For $\lambda\le\obslambda{2}$ the choice has no perceptible effect, while for
$\lambda>\obslambda{2}$ it keeps the peak at the fixed point from being split.
The histograms serve for illustration only; no transition location is extracted
from them.

\par
\textbf{Extracting the observed transition locations.}
To extract $\obslambda{1}$, we analyse the dynamics of $P^{(0)}(t)$, which is described by a sum of three complex exponentials corresponding to three poles in the Laplace domain (Supplementary Section~\ref{App:transition_locations Postselected Liouvillian with Decoherence}). Using the matrix pencil method~\cite{sarkar_using_1995}, we estimate these poles and track $e_1$, $e_2$, which form a complex-conjugate pair below the transition.
The transition occurs when this pair coalesces and splits into two distinct real values.
We extract $\obslambda{1}$ by fitting the squared pole splitting $\Delta e_{12}^2$, where $\Delta e_{12} \equiv (e_2 - e_1)/2$, near the transition to
\begin{equation}
\label{eq:approx_transition_behaviour}
    \Delta e_{12}^2 = c_1 (\lambda-\obslambda{1}),
\end{equation}
 with $c_1$ and $\obslambda{1}$ as fitting parameters.
This functional form is derived from the square-root singularity of the pole splitting $\Delta e_{12}$ characteristic of second-order exceptional points (Supplementary Section~\ref{App:extracting_critical_lambdas}), making its square $\Delta e_{12}^2$ locally linear in $\lambda$.
 Uncertainties are estimated using a bootstrap procedure. Each run is preceded by a check that the system is not in $\ket{\mathrm{B}}$, and the extracted poles are stable against truncation of $P^{(0)}(t)$ (Supplementary Section~\ref{App:extracting_QJ_transition_lambda}). 

To extract $\obslambda{2}$, we determine the critical exponent $\xi=\gamma_0/\gamma_\mathrm{a}-1$ at each $\lambda$ from two rates that set the late-time dynamics, and locate its zero crossing (Supplementary Section~\ref{App:extracting_second_transition_lambda}).
The asymptotic click rate $\gamma_0$ is the decay rate of the no-click survival probability once the trajectory has reached the fixed point, and we obtain it from $e_{3}$, the slowest pole of $P^{(0)}(t)$.
The approach rate $\gamma_\mathrm{a}$ is obtained from the conditional Bloch vector $\mathbf{R}(t)= \left( X(t),Z(t) \right)$ reconstructed by tomography, using a survival-weighted fit to its spiral towards the fixed point.
The zero crossing of $\xi(\lambda)$ is located with a quadratic fit to the ten measured values nearest the sign change.
The quoted uncertainty on $\obslambda{2}$ combines a shot-noise bootstrap of the entire analysis with systematic contributions from the fit (Supplementary Section~\ref{App:extracting_second_transition_lambda}). The dashed curve shown in Fig.~\fullref{fig:second_transition}{b} is the realistic prediction shifted by $\Delta\lambda=-0.1$ to compare its shape with the data. It is not used to determine $\obslambda{2}$.

To extract $\obslambda{4}$, we find the smallest $\lambda$ at which the no-click trajectory no longer encircles the centre of the Bloch-sphere cross-section.
From the survival-weighted fit to $\mathbf{R}(t)$ described above, we take the signed distance $R_{\min}$ of the trajectory from the centre at its first closest approach, positive when the path passes round the centre and negative otherwise.
The forbidden region opens where $R_{\min}=0$, which we locate with a cubic fit to the six measured values nearest the sign change.
The quoted uncertainty combines a shot-noise bootstrap of the entire analysis with systematic contributions from the fit (Supplementary Section~\ref{App:extracting_fr_lambda}).

To extract $\obslambda{3}$, we apply the same pole-based procedure used for $\obslambda{1}$ to the ensemble-averaged excited state probability $P_{\ket{1}}^\mathrm{ens}$ shown in Fig.~\fullref{fig:third_transition}{a}. Although $P_{\ket{1}}^\mathrm{ens}$ is described by a sum of four complex exponentials (Supplementary Section~\ref{App:transition_locations_The_Third_Transition}), numerical analysis indicates that one eigenmode decays faster than the time resolution set by the integration time $T_{\mathrm{int}}$ and is therefore not resolvable. The matrix pencil model-order selection confirms three resolvable poles in the relevant $\lambda$ range.
Using the notation introduced for the analysis of $\obslambda{1}$, we fit the squared pole splitting $\Delta e_{12}^2$ extracted from $P_{\ket{1}}^\mathrm{ens}$ to the functional form in equation~\eqref{eq:approx_transition_behaviour}, replacing $\obslambda{1}$ with $\obslambda{3}$.
Uncertainties are estimated using a bootstrap procedure. For details, see Supplementary Section~\ref{App:extracting_QZ_transition_lambda}.

\section*{Acknowledgments}
We thank Nissim Ofek for his contribution to implementing the real-time tomography experiment, and acknowledge Matan Yablon, Barak Dayan, Eran Sela, and Kater Murch for their feedback on this manuscript.
We acknowledge financial support from the Israel Science Foundation (ISF) Grant no. 2166/25. This work is part of the HQI (\href{https://www.hqi.fr}{www.hqi.fr}) and \href{https://www.lne.fr/en/metriqs-france/projet-bacq}{BACQ} initiatives and is supported by the France 2030 program under the French National Research Agency grants with numbers ANR-22-PNCQ-0002 and ANR-22-QMET-0002 (MetriQs-France).
P.K. acknowledges support from the IIT Jammu Initiation Grant No. SGT-100106.
Y.G. was supported by the Deutsche Forschungsgemeinschaft (DFG) grant SH 81/8-1 and by the National Science Foundation (NSF-2338819) and Binational Science Foundation (BSF-2023666).
S.R. is the incumbent of the Rabbi Dr. Roger Herst Career Development Chair.

\section*{Author Contributions}
B.G. and D.G. designed and performed the experiments. B.G., D.G., and N.N. analysed data and performed numerical simulations. L.M.J. performed the device fabrication. U.G. and S.H. developed code for the experiments. P.K., A.R., K.S., and Y.G. developed the theory.
B.G., D.G., N.N., K.S., A.R., P.K., Y.G., F.L., and S.R. wrote the manuscript, with input
from all authors. Y.G., F.L., K.S., and S.R. conceived and
supervised the project.

\section*{Additional information}

\noindent\textbf{Correspondence and requests for materials} should be addressed to Serge Rosenblum.

\vspace{1em}
\noindent\textbf{Competing interests:}
The authors declare no competing interests.

\vspace{1em}
\noindent\textbf{Data availability:}
The data that support the findings of this study are available from the corresponding author on request.

\vspace{1em}
\noindent\textbf{Code availability:}
The code related to the data analysis of this study is available from the corresponding author on request.

\vspace{1em}
\noindent\textbf{Use of large language models (LLMs):}
LLMs were used throughout this work. They assisted in drafting and editing the manuscript and the Supplementary Information, and in writing and debugging the code for the experiment and the data analysis. The authors reviewed, verified, and revised all text, code, and results produced with their assistance.

\clearpage

\appendix
\renewcommand{\appendixname}{SI}
\renewcommand{\thesection}{\arabic{section}}
\renewcommand{\thesubsection}{\Alph{subsection}}
\setcounter{section}{0}

\setcounter{equation}{0}
\renewcommand{\theequation}{S\arabic{equation}}
\makeatletter
\@removefromreset{equation}{section}
\makeatother

\setcounter{figure}{0}
\renewcommand{\thefigure}{S\arabic{figure}}
\renewcommand{\theHfigure}{S\arabic{figure}}

\setcounter{table}{0}
\renewcommand{\thetable}{S\arabic{table}}

\onecolumngrid           

\section{Experimental System}
\label{App:Experimental Setup and Chip Fabrication}
\begin{figure}
    \centering
    \includegraphics{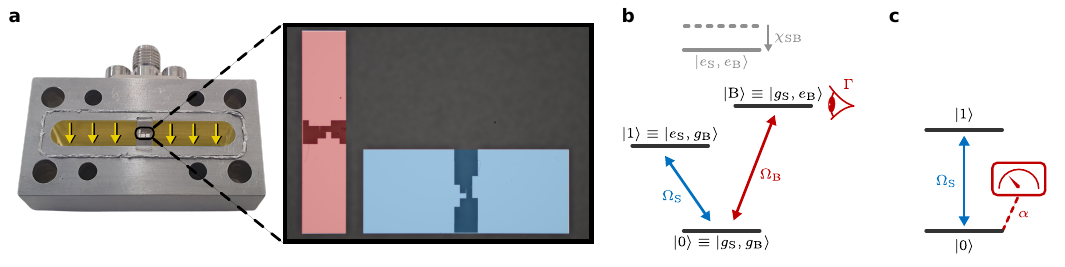}
    \caption{
    \textbf{Experimental implementation of a qubit with tunable monitoring strength.}
    \textbf{a,} Two-transmon chip mounted within a rectangular 3D superconducting cavity.
    In the expanded view, the detector (``bright'') transmon (red) is aligned with the cavity field lines (yellow arrows) to ensure strong dispersive coupling. The system (``dark'') transmon (blue) is oriented perpendicular to the field, strongly suppressing its dispersive coupling to the readout mode.
    \textbf{b,} Energy level diagram showing the shared ground state $\ket{0}$ and two single-excitation states $\ket{1}$ and $\ket{\mathrm{B}}$ forming a V-type structure. Higher-excitation states (e.g., the doubly excited state, shown in grey) lie outside the operating manifold. 
    A resonant drive of amplitude $\Omega_\mathrm{B}$ couples $\ket{0}$ to $\ket{\mathrm{B}}$, whose population is continuously monitored via the readout cavity at rate $\Gamma$.
    \textbf{c,} This configuration effectively realises a two-level system whose ground state is continuously monitored by a click detector with tunable measurement rate $\alpha \propto \Omega_\mathrm{B}^2 / \Gamma$.}
    \label{fig:cavity_chip}
\end{figure}

The device consists of two fixed-frequency transmon qubits housed in a three-dimensional superconducting readout cavity~\cite{minev_catch_2019}. The chip was fabricated on a 430-\unit{\micro\metre}-thick sapphire substrate using electron-beam lithography.
Aluminium was deposited in two orthogonal evaporations with an in situ oxidation step between them to form Manhattan-style Al/AlO$_x$/Al tunnel junctions.
The chip is placed in a rectangular $1060$ aluminium cavity (Fig.~\fullref{fig:cavity_chip}{a}), whose fundamental mode serves as the readout resonance.
The cavity halves were etched and sealed with indium wire to improve the internal quality factor.
The two transmon qubits are oriented to achieve strongly asymmetric dispersive coupling to the readout mode: the detector (``bright'') qubit is aligned parallel to the cavity field, maximising the dispersive coupling ($\chi_\mathrm{B} / 2\pi = \qty{5.1}{\MHz}$). This enables rapid, high-fidelity readout of the detector's excited state $\ket{\mathrm{B}}$. The system (``dark'') qubit is oriented perpendicularly to the cavity field, strongly suppressing its dispersive coupling ($\chi_\mathrm{S} / 2\pi = \qty{0.27}{\MHz}$). This minimises measurement-induced dephasing of the system qubit.

The nonlinear interaction between the two qubits shifts the doubly excited state $\ket{e_\mathrm{S},e_\mathrm{B}}$ by $\chi_\mathrm{SB} / 2\pi = \qty{36}{\MHz}$, rendering it inaccessible under resonant driving. This leaves an effective V-shaped three-level system (Fig.~\ref{fig:cavity_chip}{b}).

\begin{figure*}
    \centering
    \includegraphics[width=\linewidth]{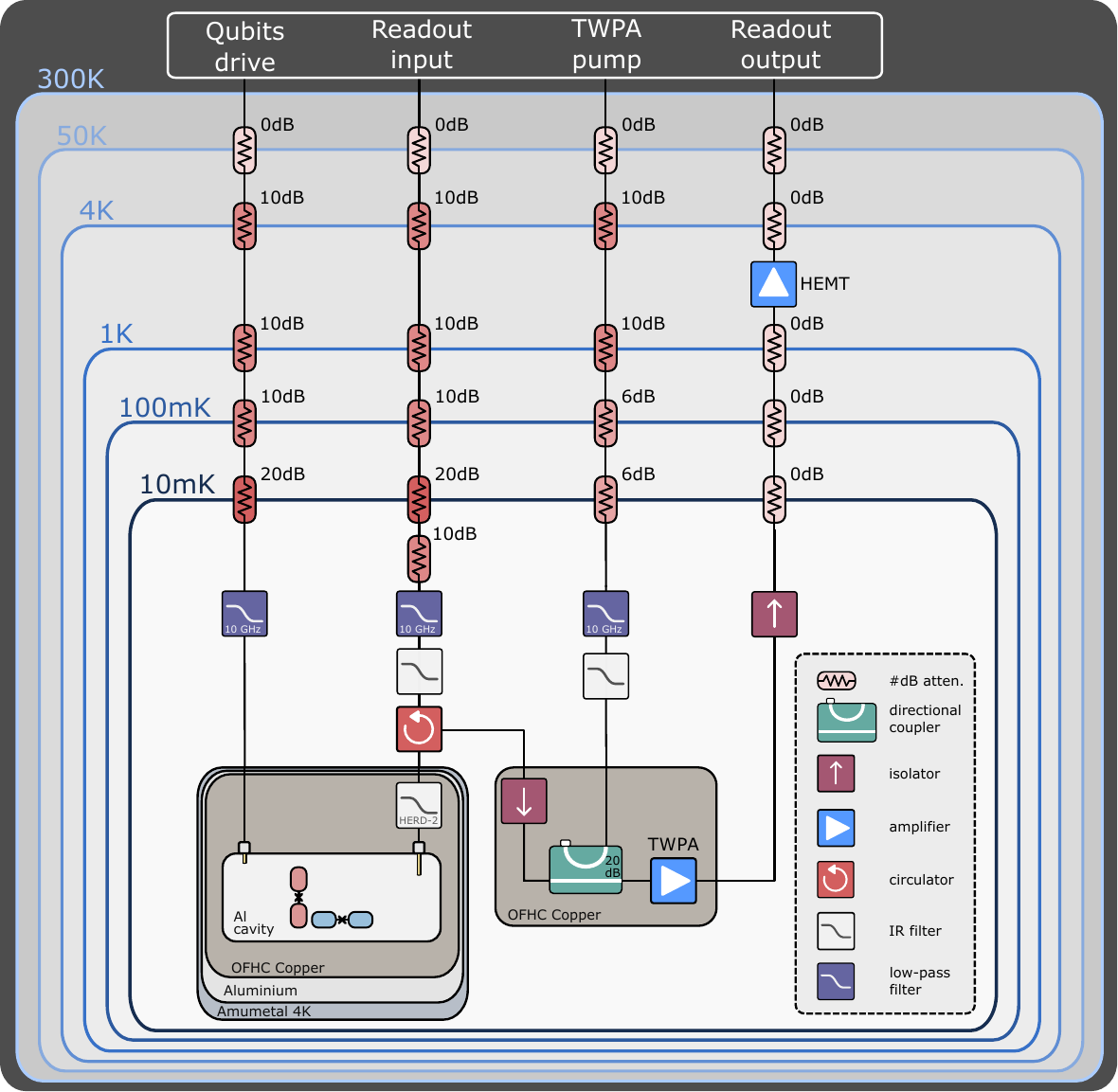}
    \caption{
    \textbf{Cryogenic microwave wiring diagram.}
    The experimental device is enclosed within OFHC copper and A4K shields for thermal and magnetic isolation.
    Control signals are synthesised by a Quantum Machines OPX+ system and upconverted using IQ-mixers before entering the cryostat.
    They pass through a sequence of attenuators, low-pass filters (LPF), and infrared filters to ensure proper thermalisation and noise suppression.
    The output signal is first amplified by a travelling-wave parametric amplifier (TWPA, Silent Waves Dreadnought) and then routed through a double-junction isolator, enabling high-fidelity single-shot readout.
    }
    \label{fig:fridge_wiring_diagram}
\end{figure*}

The cavity is enclosed within successive Amumetal and OFHC copper shields for magnetic and thermal isolation, and mounted on the mixing-chamber stage of a dilution refrigerator with base temperature of~$\sim\qty{10}{\milli\kelvin}$. The full wiring diagram is shown in Fig.~\ref{fig:fridge_wiring_diagram}.

\section{System Parameters}
\label{App:Hamiltonian and Lindbladian parameters}

The Hamiltonian describing the two transmons and readout cavity (in the absence of drives) is
\begin{equation}
\begin{split}
\hat{H}_\text{full}/ \hbar =
&\omega_\mathrm{B} \hat{b}^\dagger \hat{b}
- \frac{K_\mathrm{B}}{2} \hat{b}^{\dagger 2}\hat{b}^2
+ \omega_\mathrm{S} \hat{s}^\dagger \hat{s}
- \frac{K_\mathrm{S}}{2} \hat{s}^{\dagger 2}\hat{s}^2
\\
&- \chi_\mathrm{SB} \hat{b}^\dagger \hat{b}\, \hat{s}^\dagger \hat{s}
+ \left(
\omega_\mathrm{R}
- \chi_\mathrm{B} \hat{b}^\dagger \hat{b}
- \chi_\mathrm{S} \hat{s}^\dagger \hat{s}
\right)
\hat{r}^\dagger \hat{r},
\end{split}
\label{eq:full_hamiltonian}
\end{equation}
where $\hat{b}$, $\hat{s}$, and $\hat{r}$ are the annihilation operators for the detector transmon, system transmon, and readout cavity modes, respectively.
The parameters for this Hamiltonian, along with relevant coherence properties, are detailed in Table~\ref{Tab:HamiltonianParameters}.

\begin{table*}[h!]
    \begin{tabular}{c @{\hskip 0.5cm} l @{\hskip 0.5cm} l}
        \toprule
        \textbf{parameter} & \textbf{description} & \textbf{value} \\
        \hline
        $\omega_\mathrm{S}$ & system qubit resonance frequency & $2 \pi \times \qty{5.393}{\GHz}$ \\
        $K_\mathrm{S}$ & system qubit anharmonicity & $2 \pi \times \qty{146}{\MHz}$\\
        $T_\mathrm{S}^{1}$ & system qubit lifetime & $\qty{81}{\micro\second}$ \\
        $T_\mathrm{S}^{2}$ & system qubit coherence time & $\qty{20}{\micro\second}$ \\
        $\chi_\mathrm{S}$ & system qubit-cavity dispersive shift & $2 \pi \times \qty{0.27}{\MHz}$ \\
        $n_\mathrm{S}^\mathrm{th}$ & system qubit average thermal population & $\qty{3.5}{\%}$ \\
        $\omega_\mathrm{B}$ & detector qubit resonance frequency & $2 \pi \times \qty{6.265}{\GHz}$ \\
        $K_\mathrm{B}$ & detector qubit anharmonicity & $2 \pi \times \qty{185}{\MHz}$\\
        $T^1_\mathrm{B}$ & detector qubit lifetime ($\Omega_\mathrm{B}=0$) & $\qty{5}{\micro\second}$ \\
        $\tau_{\mathrm{B}}$ & detector-state waiting time ($\Omega_\mathrm{B}\neq0$) & $\qty{5}{\micro\second}$ \\
        $T^2_\mathrm{B}$ & detector qubit coherence time & $\qty{8}{\micro\second}$ \\
        $\chi_\mathrm{B}$ & detector qubit-cavity dispersive shift & $2 \pi \times \qty{5.1}{\MHz}$ \\
        $n_\mathrm{B}^\mathrm{th}$ & detector qubit average thermal population & $\qty{0.5}{\%}$ \\
        $\chi_\mathrm{SB}$ & system qubit-detector qubit dispersive shift & $2 \pi \times \qty{36}{\MHz}$ \\
        $\omega_\mathrm{R}$ & readout cavity resonance frequency & $2 \pi \times \qty{8.094}{\GHz}$ \\
        $\kappa_\mathrm{R}$ & readout cavity linewidth & $2 \pi \times \qty{4.6}{\MHz}$\\
        \hline \hline
    \end{tabular}
    \caption{System parameters and their respective values, cf. equation~\eqref{eq:full_hamiltonian}.
    The quoted coherence times are measured in the presence of the readout cavity drive, with the exception of $T^2_\mathrm{B}$.
    }
    \label{Tab:HamiltonianParameters}
\end{table*}

\section{Measurement Scheme}
\label{App:click_statistics}

This section describes how we continuously monitor the qubit and extract the measurement rate $\alpha$ from the recorded data.

\subsection{Drive Configuration}

Three microwave drives are applied simultaneously during the experiment (Table \ref{Tab:Pulses}):

\begin{table*}[h!]
\begin{tabular}{l @{\hskip 0.5cm} l @{\hskip 0.5cm} l @{\hskip 0.5cm} l}
\toprule
\textbf{drive} & \textbf{description} & \textbf{frequency} & \textbf{controlled parameter}
\\ \hline  & \\[-2ex]
$\Omega_\mathrm{S}$ & system qubit drive &  $\omega_\mathrm{S}$ & $\Omega_\mathrm{S}$ \\
$\Omega_\mathrm{B}$ & detector qubit drive &  $\omega_\mathrm{B}$ & $\alpha$ \\
$\Omega_\mathrm{R}$ & readout cavity drive &  $\omega_\mathrm{R} - \chi_\mathrm{B}$ & $\alpha$ \\
\hline \hline
\end{tabular}
\caption{Drives used in this work. The dimensionless measurement strength $\lambda$ is tuned by varying $\Omega_\mathrm{B}$, while $\Omega_\mathrm{S}$ and $\Omega_\mathrm{R}$ remain fixed.}
\label{Tab:Pulses}
\end{table*}

The system drive $\Omega_\mathrm{S}$ induces Rabi oscillations between $\ket{0}$ and $\ket{1}$. The detector drive $\Omega_\mathrm{B}$ couples $\ket{0}$ to the bright state $\ket{\mathrm{B}}$, and the readout drive $\Omega_\mathrm{R}$ enables continuous monitoring of the $\ket{\mathrm{B}}$ population. 

To maximise detection fidelity for \(\ket{\mathrm{B}}\) while minimising decoherence in the system manifold \(\{\ket{0},\ket{1}\}\), we drive the readout at frequency \(\omega_\mathrm{R}-\chi_\mathrm{B}\), which is resonant with the cavity only when the system is in \(\ket{\mathrm{B}}\). The resulting monitoring rate of the \(\ket{\mathrm{B}}\) population is~\cite{blais_circuit_2021}
\begin{equation}
\Gamma=\frac{\kappa_\mathrm{R}\,(\chi_\mathrm{B}/2)^2\,\bar{n}_\mathrm{B}}{2(\chi_\mathrm{B}/2)^2+(\kappa_\mathrm{R}/2)^2},
\end{equation}
where \(\kappa_\mathrm{R}\) is the readout linewidth and \(\bar{n}_\mathrm{B}\propto|\Omega_\mathrm{R}|^2\) is the average intracavity photon number.
We extract \(\bar{n}_\mathrm{B}\approx8.6\) using the calibration procedure of ref.~\cite{sank_system_2025}.

The measurement rate \(\alpha\) depends on both $\Omega_\mathrm{B}$ and $\Gamma$. The procedure for extracting $\alpha$ from experimental data is described in the final subsection below.

\subsection{Continuous Monitoring Protocol}

The readout signal is integrated over successive intervals of duration $T_{\mathrm{int}}$, producing an integrated voltage value for each interval.
\begin{figure}[h!]
    \centering \includegraphics{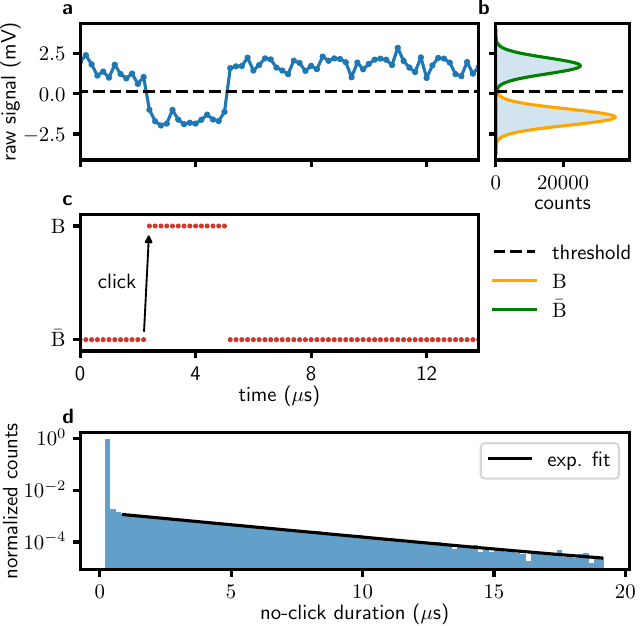}
    \caption{\textbf{Click record extraction.}
    \textbf{a,} Raw integrated measurement signal versus time.
    \textbf{b,} Histogram of measurement outcomes, fitted with a sum of two Gaussians.
    The dashed line indicates the discrimination threshold.
    \textbf{c,} Resulting binary measurement record.
    \textbf{d,} Distribution of no-click durations with $\Omega_\mathrm{S}=0$.
    The exponential fit (black line) provides an initial estimate of $\alpha$ for the hidden Markov model (HMM) analysis.
    Excess counts in the first few bins (excluded from the fit) arise from misassignments when the system is in $\ket{\mathrm{B}}$, which split a detector excursion and produce spurious short-duration no-click intervals.
    }
    \label{fig:click_record_data_analysis}
\end{figure}
These values cluster into two well-separated peaks, corresponding to the system in $\ket{\mathrm{B}}$ and in $\{\ket{0},\ket{1}\}$. We fit the histogram to a sum of two Gaussians and set the discrimination threshold at their intersection (Fig.~\fullref{fig:click_record_data_analysis}{a,b}), assigning each window to $\mathrm{B}$ or $\bar{\mathrm{B}}$.
This produces a binary record (Fig.~\fullref{fig:click_record_data_analysis}{c}).

\subsection{Definition of a Click}
In the ideal two-level model setting of ref.~\cite{snizhko_quantum_2020}, a click corresponds to instantaneous projection onto the monitored state $\ket{0}$. In our three-level system, the corresponding physical process is an \emph{excursion} to $\ket{\mathrm{B}}$: the detector drive $\Omega_{\mathrm{B}}$, together with continuous readout of $\ket{\mathrm{B}}$, induces incoherent transitions from $\ket{0}$ to $\ket{\mathrm{B}}$ at rate $\alpha$, and the system returns to $\ket{0}$ after a mean waiting time $\tau_\mathrm{B}$. The net effect of an excursion is projection onto $\ket{0}$, as in the ideal model, but with a finite average delay $\tau_\mathrm{B}$. The impact of this delay on the dynamics is analysed in Section~\ref{App:realistic_model}.
Since $\tau_\mathrm{B}\gg T_\mathrm{int}$, an excursion typically spans several consecutive windows assigned to $\mathrm{B}$. We define a \emph{click} as a $\mathrm{B}$-assigned window following a $\bar{\mathrm{B}}$-assigned window. Clicks terminate no-click intervals and define $P^{(0)}(t)$ and the conditional dynamics. An excursion may fail to produce a click (false negative), and a misassigned window may produce a click without an excursion (false positive; Supplementary Section~\ref{App:detector_inefficiency_realistic_model}).

\subsection{Extraction of the Measurement Rate\texorpdfstring{ $\alpha$}{}}
\label{App:alpha_extraction}

From the binary measurement record, we compute histograms of no-click durations, defined as intervals between successive registered clicks. With the system drive off ($\Omega_\mathrm{S}=0$) and the system initialised in $\ket{0}$, these distributions decay exponentially at a rate close to $\alpha(\Omega_\mathrm{B})$,
consistent with a Poisson process (Fig.~\fullref{fig:click_record_data_analysis}{d}).
The exponential fit provides an initial estimate of $\alpha$, but is affected by finite detection resolution and readout errors.
To obtain more accurate values, we analyse the binary measurement record using a three-state hidden Markov model (HMM)~\cite{baum_maximization_1970} that accounts for misassignments.
This HMM is defined by a transition matrix~$\mathbf{T}$ and an emission matrix~$\mathbf{E}$. The transition matrix encodes the probabilities $p_{ij}$ of moving from state $\ket{i}$ to state $\ket{j}$ in one time step~$dt$. In our system, transitions occur between $\ket{0}$ and $\ket{\mathrm{B}}$ and between $\ket{0}$ and $\ket{1}$, but not directly between $\ket{\mathrm{B}}$ and $\ket{1}$. The matrix therefore has the form:

\begin{equation}
\mathbf{T} = \begin{pmatrix}
1 - p_{0\mathrm{B}} - p_{01} & p_{0\mathrm{B}} & p_{01} \\
p_{\mathrm{B}0} & 1 - p_{\mathrm{B}0} & 0 \\
p_{10} & 0 & 1 - p_{10}
\end{pmatrix},
\end{equation}
where rows and columns are ordered as $\ket{0}$, $\ket{\mathrm{B}}$, $\ket{1}$. 
These probabilities relate to the underlying rates via
\[
\left( p_{0j},\,p_{j0} \right) \;=\;
\Bigg(
\frac{\gamma_{j\uparrow}}{\Gamma_j}\bigl(1-e^{-\Gamma_j dt}\bigr),\;
\frac{\gamma_{j\downarrow}}{\Gamma_j}\bigl(1-e^{-\Gamma_j dt}\bigr)
\Bigg),
\quad \Gamma_j=\gamma_{j\uparrow}+\gamma_{j\downarrow},
\]

\noindent for $j \in \{\mathrm{B}, 1\}$, where $\gamma_{j\uparrow}$ and $\gamma_{j\downarrow}$ are the excitation and relaxation rates between $\ket{0}$ and $\ket{j}$.
These relate to the parameters used elsewhere in this paper as follows: the measurement rate is $\alpha = \gamma_{\mathrm{B}\uparrow}$, the inverse waiting time in $\ket{\mathrm{B}}$ is $\tau_\mathrm{B}^{-1} = \gamma_{\mathrm{B}\downarrow}$, and the qubit relaxation rate is $(T_\mathrm{S}^1)^{-1} = \gamma_{1\downarrow}$.

 The emission matrix encodes the probability of each assignment given the hidden state. Only $\ket{\mathrm{B}}$ should be assigned to $\mathrm{B}$, while $\ket{0}$ and $\ket{1}$ should both be assigned to $\bar{\mathrm{B}}$. In practice, the readout occasionally misassigns an integration window. We write $p(\mathrm{B}|\bar{\mathrm{B}})$ for the probability of a spurious $\mathrm{B}$ assignment (system in $\ket{0}$ or $\ket{1}$) and $p(\bar{\mathrm{B}}|\mathrm{B})$ for the probability of a missed $\mathrm{B}$ assignment (system in $\ket{\mathrm{B}}$).
 Both are properties of the readout of a single integration window, as illustrated in Fig.~\fullref{fig:click_record_data_analysis}{b}.
 The emission matrix is therefore
\begin{equation}
    \mathbf{E} =
\begin{pmatrix}
1 - p(\mathrm{B}|\bar{\mathrm{B}}) & p(\mathrm{B}|\bar{\mathrm{B}}) \\[4pt]
p(\bar{\mathrm{B}}|\mathrm{B}) & 1 - p(\bar{\mathrm{B}}|\mathrm{B}) \\[4pt]
1 - p(\mathrm{B}|\bar{\mathrm{B}}) & p(\mathrm{B}|\bar{\mathrm{B}})
\end{pmatrix},
\end{equation}
with columns ordered as $\bar{\mathrm{B}}$, $\mathrm{B}$. The identical first and third rows reflect that $\ket{0}$ and $\ket{1}$ are indistinguishable to the detector.

\begin{figure}
    \centering
    \includegraphics[]{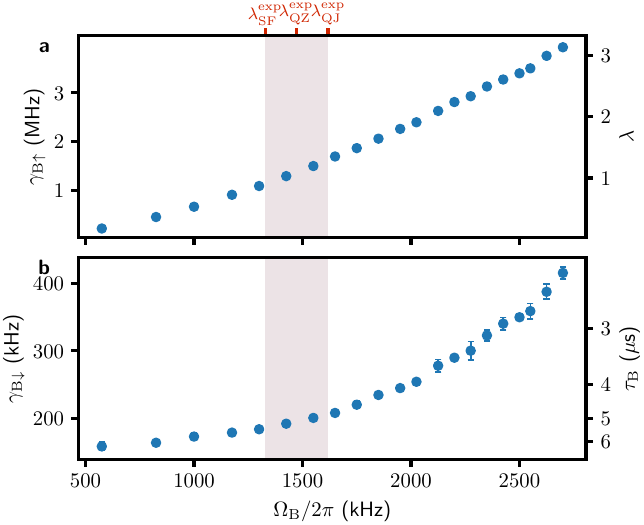}
\caption{\textbf{Extraction of measurement parameters via HMM analysis.}
\textbf{a,}~Measurement rate $\alpha = \gamma_{\mathrm{B}\uparrow}$ versus detector drive amplitude $\Omega_\mathrm{B}$, with $\lambda = \alpha/(2\Omega_\mathrm{S})$ on the right axis.
Red ticks on the top axis mark the drive amplitudes at which $\lambda$ reaches the observed transition values $\obslambda{2}$, $\obslambda{3}$ and $\obslambda{1}$. The shaded band highlights the transition region in both panels.
\textbf{b,}~Decay rate $\gamma_{\mathrm{B}\downarrow} = \tau_\mathrm{B}^{-1}$ versus $\Omega_\mathrm{B}$, with $\tau_\mathrm{B}$ on the right axis. $\gamma_{\mathrm{B}\downarrow}$ is less affected by the drive owing to the Stark shift induced by readout photons. Each point is the weighted mean of four repeated measurements; error bars are one standard error, including the spread between repetitions, and are smaller than the markers where not visible.}
\label{fig:hmm_calib_with_error}
\end{figure}

We use the Baum--Welch algorithm~\cite{baum_maximization_1970} to find the parameters that maximise the likelihood of the observed click record. Initial guesses for $\gamma_{1\uparrow}$ and $\gamma_{1\downarrow}$ come from independent $T^1_\mathrm{S}$ and thermal population measurements, those for $\gamma_{\mathrm{B}\uparrow}$ and $\gamma_{\mathrm{B}\downarrow}$ from exponential fits to the durations of consecutive $\mathrm{B}$- and $\bar{\mathrm{B}}$-assigned windows, and those for $p(\mathrm{B}|\bar{\mathrm{B}})$ and $p(\bar{\mathrm{B}}|\mathrm{B})$ from the Gaussian separation in the readout histogram (see Fig.~\ref{fig:click_record_data_analysis}).
From the optimised transition matrix, we extract $\alpha$ and $\tau_\mathrm{B}$, while the emission matrix provides the readout misassignment probabilities $p(\mathrm{B}|\bar{\mathrm{B}})$ and $p(\bar{\mathrm{B}}|\mathrm{B})$.
These experimentally extracted parameters are used in the numerical
simulations, with the additional contribution to the false-negative
probability from missed short excursions accounted for separately (Supplementary Sections~\ref{App:realistic_model} and~\ref{App:numerical_sim}).

Figure~\ref{fig:hmm_calib_with_error} shows the extracted rates versus detector drive amplitude $\Omega_\mathrm{B}$. The measurement rate $\alpha$ increases approximately quadratically with $\Omega_\mathrm{B}$, consistent with the expected scaling $\alpha \propto \Omega_\mathrm{B}^2/\Gamma$. In contrast, the waiting time $\tau_\mathrm{B}$ varies less across the range of drive amplitudes. This is because the readout photons induce a Stark shift that detunes the detector qubit transition once the system reaches $\ket{\mathrm{B}}$. Instead, $\tau_\mathrm{B}$ is determined primarily by the intrinsic lifetime of the detector qubit $T^1_\mathrm{B}\approx \qty{5}{\micro\second}$. Indeed, in the range of $\Omega_\mathrm{B}$ where the dynamical transitions occur, we find $\tau_\mathrm{B} \approx \qty{5}{\micro\second}$ (Supplementary Table~\ref{Tab:HamiltonianParameters}).

\section{Overview of the Ideal Model of Monitored Qubit Dynamics}
\label{App:ideal_model}

In this section, we review the ideal monitored-qubit model of
ref.~\cite{snizhko_quantum_2020} in the conventions of our experiment.
We monitor $\ket{0}$ rather than $\ket{1}$, and write the drive along
$\hat{\sigma}_\mathrm{y}$ rather than $\hat{\sigma}_\mathrm{x}$, with
$\Omega_{\mathrm S}=2\Omega_{\mathrm s}$ in the notation of that work.
We summarise its no-click dynamics, steady-state angular distribution,
and ideal transition locations, then introduce the angular dwell-time
density used in our experiment.

\subsection{No-Click Dynamics of a Monitored Qubit}

We consider a coherently driven qubit with Hamiltonian 
\begin{equation}
\label{eq:drive_hamiltonian}
\hat{H}_\mathrm{d} / \hbar = \frac{\Omega_\mathrm{S}}{2}\hat{\sigma}_\mathrm{y}.
\end{equation}
The qubit ground state $\ket{0}$ is monitored by a sequence of partial measurements performed at intervals~$dt\ll\Omega_\mathrm{S}^{-1}$. 
Each measurement yields either a \emph{click} or \emph{no-click} outcome, associated with the Kraus operators
\begin{equation}
\label{eq:backaction_ideal_case}
\hat{M}_{\text{click}} = \sqrt{p}\,\ketbra{0}{0},
\qquad
\hat{M}_{\text{no-click}} = \ketbra{1}{1} + \sqrt{1 - p}\,\ketbra{0}{0},
\end{equation}
with $p\in[0,1]$ the per-interval click probability.
In the limit $dt \! \to \! 0$, we keep the effective measurement rate $\alpha \equiv \frac{p}{dt}$
fixed, so that $p=\alpha\,dt\ll 1$ per time step.
The dimensionless measurement strength is then $\lambda = \alpha/({2\Omega_\mathrm{S}})$.

Expanding $\hat{M}_{\text{no-click}}$ to first order in $p$ and combining with the unitary evolution $\hat{U} \approx \mathbb{I} - i\frac{\Omega_\mathrm{S}}{2}\hat{\sigma}_\mathrm{y} \, dt$, the no-click evolution takes the form $\hat{M}_{\text{no-click}}\hat{U} \approx \mathbb{I} - \frac{i}{\hbar}\hat{H}_{\mathrm{eff}}\,dt$, yielding the effective non-Hermitian Hamiltonian of equation~\eqref{Eq:non-hermitian hamiltonian}:
\begin{equation}
    \hat{H}_{\mathrm{eff}}/\hbar = \frac{\Omega_\mathrm{S}}{2}\left(\hat{\sigma}_\mathrm{y} - 2i\lambda\,\ketbra{0}{0}\right).
\end{equation}

Since the dynamics are constrained to the $XZ$-plane for a qubit initialised in either \(\ket{0}\) or \(\ket{1}\), the qubit state is fully characterised by the polar angle $\theta(t)$ of its Bloch vector $\ket{\psi[\theta(t)]}=\cos \!\left(\frac{\theta(t)}{2}\right)\ket{1}+ \sin \!\left(\frac{\theta(t)}{2}\right)\ket{0}$.
Here \(\theta=0\) is the unmonitored state \(\ket{1}\), while \(\theta=\pi\) is the monitored state \(\ket{0}\).
The combined evolution leads to the equation of motion~\cite{snizhko_quantum_2020}:
\begin{equation}
\label{eq:theta_of_t_eq_of_motion}
\theta(t+dt) =
\begin{cases}
\theta(t) - \Omega_{\mathrm{S}}\,[1 + \lambda \sin \theta(t)]\,dt & \text{no-click} \\[6pt]
\pi & \text{click},
\end{cases}
\end{equation}
where the click probability per interval is 
\begin{equation}
\label{eq:click_prob_at_each_infinitesimal_step}
\begin{split}
    p_{\text{click}}\left(\theta\right) &= \alpha dt \,\sin^2 \!\left(\frac{\theta}{2}\right) \\
    &\equiv r\left(\theta\right)dt, 
\end{split}
\end{equation}
with $r\left(\theta\right)$ the instantaneous click rate.

The no-click dynamics depend on $\lambda$: for $\lambda < 1$, the angular velocity is always negative, driving $\theta(t)$ anticlockwise around the Bloch sphere. For $\lambda > 1$, two fixed points emerge, $\theta_{\pm} = 2 \arctan \!\left(-\lambda \pm \sqrt{\lambda^2 - 1}\right)$, where the angular velocity vanishes. In this regime, the angular velocity is positive for $\theta_{-} < \theta < \theta_{+}$, reflecting the competition between the coherent drive and the measurement backaction.
Consequently, no-click trajectories starting at $\theta=0$ or
$\theta=\pi$ cannot enter the \emph{forbidden region} $\theta\in(-\pi,\theta_+]$. This includes trajectories reset to
$\theta=\pi$ by a detector click.
Solving equation~\eqref{eq:theta_of_t_eq_of_motion} as in
ref.~\cite{snizhko_quantum_2020} (Appendix B), but with the initial
condition $\theta(0)=\pi$ corresponding to our monitored state
$\ket{0}$, gives
\begin{equation}
    \label{eq:theta_t_solution_initial_theta_pi}
    \tan \frac{\theta(t)}{2} = 
    \frac{\sqrt{1-\lambda^2}}{\tan \!\left(\tfrac{1}{2}\Omega_{\mathrm{S}} t \sqrt{1-\lambda^2}\,\right)} - \lambda.
\end{equation}

For $\lambda < 1$, this describes periodic evolution,
while for $\lambda > 1$, the qubit evolves from $\theta=\pi$ at $t=0$ to $\theta=\theta_{+}$ at $t=\infty$ --- a continuous quantum jump.
The transition between these regimes, nominally at $\critlambda{1}=1$, is visible in Fig.~\fullref{fig:first_transition}{a--b}.

\subsection{Click Record Statistics}
We now turn to the no-click survival probability $P^{(0)}(t)$, defined as the probability of obtaining no clicks up to time $t$.
Its evolution is governed by
\begin{equation}
\label{eq:dP0_dt_p_over_dt_times_P0_t}
\frac{d P^{(0)}(t)}{dt} = -r\left(\theta(t)\right)\,P^{(0)}(t),
\end{equation}
where $-dP^{(0)}(t)/dt$ corresponds to the probability density of registering the \emph{first} click at time~$t$, or, equivalently, the distribution of no-click durations.

Combining equation~\eqref{eq:click_prob_at_each_infinitesimal_step} with the trajectory $\theta(t)$ from equation~\eqref{eq:theta_t_solution_initial_theta_pi} yields
\begin{equation}
\label{eq:P^0(t)_initial_theta_pi}
P^{(0)}(t) = e^{- \lambda \Omega_{\mathrm{S}} t} \frac{
-1
+ \lambda^2 \cos \left(\Omega_{\mathrm{S}} t \sqrt{1-\lambda^2}\right)
+\lambda \sqrt{1-\lambda^2} \sin \left(\Omega_{\mathrm{S}} t \sqrt{1-\lambda^2}\right)
}{\lambda^2-1}.
\end{equation}

Consequently, the distribution of no-click durations is

\begin{equation}
\label{eq:-dP0_dt_initial_theta_pi}
-\frac{d P^{(0)}(t)}{d t}=\lambda \Omega_{\mathrm{S}} e^{- \lambda \Omega_{\mathrm{S}} t} \frac{
-1
+ \left(2 \lambda^2-1\right) \cos \left(\Omega_{\mathrm{S}} t \sqrt{1-\lambda^2}\right)
+ 2 \lambda \sqrt{1-\lambda^2} \sin \left(\Omega_{\mathrm{S}} t \sqrt{1-\lambda^2}\right)
}{\lambda^2-1}.
\end{equation}

For $\lambda<1$,  the first-click density $-dP^{(0)}(t)/dt$ has an oscillatory modulation, reflecting the periodic no-click trajectory: since $r(\theta)=\alpha\sin^2(\theta/2)$ (equation~\eqref{eq:click_prob_at_each_infinitesimal_step}), clicks cluster near $\theta=\pi$ and are suppressed near $\theta=0$. For $\lambda>1$, the oscillations cease, and the distribution of no-click durations decays exponentially, cf. Fig.~\fullref{fig:first_transition}{c}.
$P^{(0)}(t)$ exhibits the same regimes (oscillating for $\lambda < 1$ and non-oscillating for $\lambda > 1$), making it a suitable observable for identifying the location of the quantum-jump transition, cf. Fig.~\fullref{fig:crit_lambda_1_with_poles}.

The expressions above start at $\theta(\ket{0})=\pi$, whereas the
experimental no-click traces start at $\theta(\ket{1})=0$.
Their time dependence differs, but the no-click eigenvalues and the ideal transition at $\lambda=1$ are the same for both initial states.

\subsection{Angular Distributions \label{App:angular distributions}}
For $\lambda>1$, ref.~\cite{snizhko_quantum_2020}
characterises state freezing through the steady-state angular
probability density $P_\infty(\theta)$: 
\begin{equation}
    \label{eq:P_inf(theta)_for_lambda_greater_than_1}
    P_\infty\left(\theta\right)=\begin{cases}
         \frac{\lambda}{(1+\lambda \sin \theta)^2}
         \left(
         \frac{\tan \frac{\theta}{2}-\tan \frac{\theta_{+}}{2}} {\tan \frac{\theta}{2}-
         \cot \frac{\theta_{+}}{2} }
         \right)^{\frac{\lambda}{\sqrt{\lambda^2-1}}} & \text{for } \theta \in\left(\theta_{+} , \pi\right]
         \\[6pt]
         0 & \text{otherwise}.
    \end{cases}
\end{equation}
This expression diverges at $\theta_+$ for $\lambda > 2/\sqrt{3}$, indicating that the system spends an extended amount of time in the vicinity of $\theta_+$.

Here, we instead characterise the state-freezing transition using the angular dwell-time density $\tau_\theta(\theta)$. Consider a system initialised at $\theta = \pi$. We define the expectation value $\mathbb{E}[T_{\Delta\theta}(\theta)]$ of the time $T_{\Delta\theta}(\theta)$ spent in the interval $[\theta,\, \theta + \Delta\theta)$ before a click occurs. The dwell-time density is then
\begin{equation}
\label{eq:tau_theta_definition}
\tau_{\theta}(\theta) \equiv
\lim_{\Delta\theta\to 0}\frac{\mathbb{E}\left[T_{\Delta\theta}(\theta)\right]}{\Delta\theta}.
\end{equation}
Since clicks reset the system to $\theta = \pi$, one expects that $P_\infty(\theta) \propto \tau_{\theta}(\theta)$. This reflects the fact that the probability density of finding the system state in the vicinity of $\theta$ after an infinite evolution duration is proportional to the time the system spends in that vicinity. Equation \eqref{eq:tau_theta_explicit_expression} below confirms that this is indeed the case.

The expectation value $\mathbb{E}\left[T_{\Delta\theta}(\theta)\right]$ can be calculated as follows. For small $\Delta\theta$ and conditioned on the absence of clicks, the time spent in the interval $[\theta,\, \theta + \Delta\theta)$ is simply the traversal time $T_{\Delta\theta}(\theta) = \Delta\theta/|\dot{\theta}|$, where the angular velocity satisfies $\dot{\theta}(\theta) < 0$, cf.~equation~\eqref{eq:theta_of_t_eq_of_motion}. However, when starting from $\theta = \pi$, the system does not always reach this interval: a click may interrupt its trajectory, in which case $T_{\Delta\theta}(\theta) = 0$. The probability of reaching $\theta$ is $P^{(0)}(\theta)\equiv P^{(0)}(t(\theta))$, where $t(\theta)$ is the time to reach $\theta$ under no-click evolution. Thus, the system spends on average $\mathbb{E}\left[T_{\Delta\theta}(\theta)\right] = P^{(0)}(\theta) \Delta\theta/|\dot{\theta}|$ in the interval. Substituting into equation~\eqref{eq:tau_theta_definition} yields
\begin{equation}
\label{eq:tau_theta_P0_theta_over_theta_dot}
\tau_{\theta}(\theta) = \frac{P^{(0)}(\theta)}{|\dot{\theta}|}.
\end{equation}

The dwell time can also be expressed in terms of the angular first-click density $f_\theta(\theta)$, defined as the probability density for the first click to occur when the system reaches $\theta$:
\begin{equation}
\label{eq:rho_definition_supplemental}
    f_\theta(\theta)\equiv \frac{d P^{(0)}(\theta)}{d \theta} = -\frac{dP^{(0)}(t)}{dt} \frac{1}{|\dot{\theta}|}.
\end{equation}
Using equation~\eqref{eq:dP0_dt_p_over_dt_times_P0_t}, we obtain
\begin{equation}f_\theta\left(\theta\right) = r\left(\theta\right) \frac{P^{(0)} \left(\theta\right)}{|\dot{\theta}|},
\end{equation}
which yields 
\begin{equation}
\label{eq:rho_theta_p0_theta_equivalence}  \tau_{\theta}(\theta) =\frac{P^{(0)} \! \left(\theta\right)}{\left| \dot{\theta} (\theta)\right|}=\frac{f_\theta\left(\theta\right)}{r\left(\theta\right)}.
\end{equation}

Using the survival probability and no-click trajectory derived above,
we obtain the angular first-click density for a trajectory starting at
$\theta=\pi$:
\begin{equation}
\label{eq:rho(theta)}
    f_\theta\left(\theta\right)=
    \begin{cases}
         \frac{2\lambda\sin^2 \! {\left( \frac{\theta}{2} \right)}}{\left(1+\lambda \sin \theta\right)^2}
    \left( \frac{\tan \frac{\theta}{2} - \tan \frac{\theta_{+}}{2}}
    {\tan \frac{\theta}{2} - \cot \frac{\theta_{+}}{2}} \right)^{\frac{\lambda}{\sqrt{\lambda^2 - 1}}}
    & \text{for }\theta \in\left(\theta_{+} , \pi\right]
         \\[6pt]
         0 & \text{otherwise}.
    \end{cases}
\end{equation}
This derivation adapts the angular survival calculation of
ref.~\cite{snizhko_quantum_2020} (Appendix B), which instead starts
at $\theta=0$.
Substituting this result into equation~\eqref{eq:rho_theta_p0_theta_equivalence} gives
\begin{equation}
    \label{eq:tau_theta_explicit_expression}
    \tau_{\theta}(\theta)
    =  \begin{cases}
         \frac{1}{\Omega_\mathrm{S}\left(1+\lambda \sin \theta\right)^2}
    \left( \frac{\tan \frac{\theta}{2} - \tan \frac{\theta_{+}}{2}}
    {\tan \frac{\theta}{2} - \cot \frac{\theta_{+}}{2}} \right)^{\frac{\lambda}{\sqrt{\lambda^2 - 1}}}
     & \text{for }\theta \in\left(\theta_{+} , \pi\right]
         \\[6pt]
         0 & \text{otherwise}.
    \end{cases}
\end{equation}
Thus, for trajectories starting in the monitored state immediately
after a click,
$\tau_\theta(\theta)=P_\infty(\theta)/(\lambda\Omega_{\mathrm S})$.

This expression can be rewritten as
\begin{equation}
    \label{eq:tau_theta_explicit_practical}
    \tau_{\theta}(\theta)
    =  \begin{cases}
         \frac{1}{\Omega_\mathrm{S}} \frac{1}{\cos^4\frac{\theta}{2}} 
    \left( \tan \frac{\theta}{2} - \tan \frac{\theta_{+}}{2}\right)^{\frac{\lambda}{\sqrt{\lambda^2 - 1}} - 2} \left( \tan \frac{\theta}{2} - \tan \frac{\theta_{-}}{2}\right)^{-\frac{\lambda}{\sqrt{\lambda^2 - 1}} - 2} & \text{for }\theta \in\left(\theta_{+} , \pi\right]
         \\[6pt]
         0 & \text{otherwise}.
    \end{cases}
\end{equation}
Here we used the relation
\begin{equation}
1 + \lambda \sin \theta = \cos^2\frac{\theta}{2} \left(\tan\frac{\theta}{2} - \tan\frac{\theta_+}{2}\right) \left(\tan\frac{\theta}{2} - \tan\frac{\theta_-}{2}\right).
\end{equation}
The divergence in $\tau_{\theta}(\theta)$ thus arises when the critical exponent
$\xi(\lambda)\equiv \left(\lambda/\sqrt{\lambda^2-1}\right) -2$ crosses $0$, i.e., at
$\lambda=\critlambda{2}=2/\sqrt{3}$.

When initialising the system in $\theta = 0$ (as in the main text), rather than at $\theta = \pi$ (as above), there are two differences: (i) the system does not visit $\theta > 0$ before a click occurs; (ii) $\tau_{\theta}(\theta)$ is multiplied by a constant factor because in the new setting $P^{(0)}(\theta = 0) = 1$ (as opposed to $P^{(0)}(\theta = \pi) = 1$).

\section{Realistic Model of Monitored Qubit Dynamics}
\label{App:realistic_model}

In this section, we extend the ideal model of Supplementary Section~\ref{App:ideal_model} and ref.~\cite{snizhko_quantum_2020} to account for realistic experimental conditions:

\begin{itemize}
    \item Pure dephasing, modelled as Markovian fluctuations of the relative phase between $\ket{0}$ and $\ket{1}$, quantified by the pure dephasing time $T^\phi_\mathrm{S}$. This dephasing is primarily induced by the readout cavity drive;
    \item Relaxation, describing spontaneous decay from $\ket{1}$ to $\ket{0}$ with lifetime $T^1_\mathrm{S}$;
     \item Finite waiting time $\tau_\mathrm{B}$ in the detector qubit's excited state $\ket{\mathrm{B}}$ (cf.~Fig.~\ref{fig:hmm_calib_with_error}): during each excursion, the system remains in $\ket{\mathrm{B}}$ for a finite time before relaxing back to $\ket{0}$;
    \item Finite integration time $T_{\mathrm{int}}$, arising because the detection signal is integrated over a nonzero window, reducing temporal resolution compared to the idealised instantaneous measurements.
    \item Detector inefficiency, manifesting as false positives (clicks without an excursion, at rate $\kappa_\mathrm{FP}\equiv p(\mathrm{B}|\bar{\mathrm{B}})/T_{\mathrm{int}}$, cf. Supplementary Section~\ref{App:alpha_extraction}) and false negatives (excursions that produce no click, with probability $p_\mathrm{FN}$);

\end{itemize}

Dephasing, relaxation, detector inefficiency, and finite waiting time admit analytical treatment, developed below. Finite integration time enters the analytical estimate of the false-negative probability and is incorporated into the numerical simulations (Section~\ref{App:numerical_sim}).

\subsection{Dephasing and Relaxation}

The pure-state dynamics of Section~\ref{App:ideal_model} assumed ideal measurements and no decoherence. To incorporate dephasing and relaxation, we use the density matrix formalism.

\paragraph{Ensemble dynamics.}
The unconditional evolution is governed by the Lindblad master equation~\eqref{Eq:full_liouvillian}, extended to include dephasing and relaxation:
\begin{equation}
\label{eq:full_lindblad}
\frac{d\hat{\rho}}{dt} = \mathcal{L}[\hat{\rho}] = -\frac{i}{\hbar}[\hat{H}_\mathrm{d},\hat{\rho}] + \alpha\,\mathcal{D}[\ketbra{0}{0}]\hat{\rho} + \frac{1}{2T^\phi_\mathrm{S}}\mathcal{D}[\hat{\sigma}_\mathrm{z}]\hat{\rho} + \frac{1}{T^1_\mathrm{S}}\mathcal{D}[\ketbra{0}{1}]\hat{\rho},
\end{equation}
where $\mathcal{D}[\hat{L}]\hat{\rho} = \hat{L}\hat{\rho}\hat{L}^\dagger - \frac{1}{2}\{\hat{L}^\dagger\hat{L}, \hat{\rho}\}$ is the standard dissipator. The three dissipative terms describe the measurement backaction, dephasing, and relaxation, respectively.
In particular, the residual dispersive coupling to the readout cavity allows it to weakly distinguish $\ket{0}$ from $\ket{1}$. Because this information is not resolved in the binary detector record, the associated backaction appears predominantly as unobserved dephasing. This contribution is included in $1/T^\phi_\mathrm{S}$, determined from coherence measurements with the readout drive present (Table~\ref{Tab:HamiltonianParameters}).

\paragraph{Postselected dynamics.}
For evolution conditioned on no clicks, the reset term $\hat{L}\hat{\rho}\hat{L}^\dagger$ of the measurement dissipator is removed, yielding the postselected Liouvillian $\mathcal{L}_\mathrm{P}$:
\begin{equation}
\label{eq:postselected_lindblad}
\frac{d\hat{\rho}}{dt} = \mathcal{L}_\mathrm{P}[\hat{\rho}] = -\frac{i}{\hbar}[\hat{H}_\mathrm{d},\hat{\rho}] - \frac{\alpha}{2}\{\ketbra{0}{0}, \hat{\rho}\} + \frac{1}{2T^\phi_\mathrm{S}}\mathcal{D}[\hat{\sigma}_\mathrm{z}]\hat{\rho} + \frac{1}{T_\mathrm{S}^1}\mathcal{D}[\ketbra{0}{1}]\hat{\rho}.
\end{equation}
This evolution is not trace-preserving: $\mathrm{Tr}[\hat{\rho}(t)]$ gives the no-click survival probability $P^{(0)}(t)$. In the absence of decoherence, equation~\eqref{eq:postselected_lindblad} reduces to the pure-state no-click dynamics of Section~\ref{App:ideal_model}.
Pure dephasing satisfies $\mathcal{D}[\hat{\sigma}_\mathrm{z}]=4\mathcal{D}[\ketbra{0}{0}]=4\mathcal{D}[\ketbra{1}{1}]$, so its effect is independent of which state projector is used to represent it. Crucially, conditioning on no detector clicks does not condition on this unobserved channel: the full dephasing dissipator remains in the no-click evolution. Additional dephasing is therefore not equivalent to increasing
the monitored measurement rate $\alpha$.
In Section~\ref{App:transition_locations} we show how dephasing affects the different transition locations in contrasting ways.

\subsection{Finite waiting time in \texorpdfstring{$\ket{\mathrm{B}}$}{B}}

During an excursion, the system is projected onto $\ket{\mathrm{B}}$ and returns to the qubit subspace only after a finite waiting time $\tau_\mathrm{B}$, modifying the ensemble dynamics:
\begin{align}
    \partial_t \hat{\rho}_\mathrm{S}(t) &= \mathcal{L}_\mathrm{P}[\hat{\rho}_\mathrm{S}] + \frac{p_\mathrm{B}}{\tau_\mathrm{B}} \ketbra{0}{0},  \nonumber\\
    \partial_t p_\mathrm{B}(t) &= \alpha \bra{0} \hat{\rho}_\mathrm{S} \ket{0} - \frac{p_\mathrm{B}}{\tau_\mathrm{B}}, \label{eq:S_evolution}
\end{align}
where $\hat{\rho}_\mathrm{S}$ is the density matrix in the qubit subspace and $p_\mathrm{B}$ is the population in $\ket{\mathrm{B}}$. Since the ensemble dynamics are not influenced by detection errors, this expression uses $\mathcal{L}_\mathrm{P}$ rather than $\widetilde{\mathcal{L}}_\mathrm{P}$. In the limit $\tau_\mathrm{B}\to \infty$, the system never returns from $\ket{\mathrm{B}}$, and the qubit dynamics reduce to the postselected evolution. In the opposite limit $\tau_\mathrm{B} \to 0$, the population returns instantaneously: setting $\partial_t p_\mathrm{B} = 0$ gives $p_\mathrm{B}/\tau_\mathrm{B} = \alpha\bra{0}\hat{\rho}_\mathrm{S}\ket{0}$, and substituting into the first equation recovers the ensemble dynamics of equation~\eqref{eq:full_lindblad}.
The indirect effect of the waiting time on excursions that produce no click is considered in the following section.

\subsection{Detector Inefficiency}
\label{App:detector_inefficiency_realistic_model}

In practice, the measurement record is affected by detection errors: false positives (a click without an excursion) and false negatives (an excursion that produces no click).
False positives occur at a rate $\kappa_\mathrm{FP}$, assumed to be independent of the state within the qubit manifold $\{\ket{0},\ket{1}\}$. Using the value extracted from the HMM analysis (Supplementary Section~\ref{App:alpha_extraction}), $p(\mathrm{B}|\bar{\mathrm{B}})=\qty{0.5}{\percent}$, we obtain $\kappa_\mathrm{FP}=p(\mathrm{B}|\bar{\mathrm{B}})/T_{\mathrm{int}}=\qty{2.5e4}{\per\second}$.
False negatives arise predominantly from short excursions to $\ket{\mathrm{B}}$ that are not resolved within the finite integration time $T_{\mathrm{int}}$. The intrinsic probability $p(\bar{\mathrm{B}}|\mathrm{B})$ of misassigning a window in which $\ket{\mathrm{B}}$ is occupied throughout is \qty{0.5}{\percent} (HMM analysis). Because an excursion spans $\tau_\mathrm{B}/T_{\mathrm{int}}\approx 25$ windows on average, it is missed only if every window is misassigned. This contributes less than $10^{-3}$ to $p_\mathrm{FN}$ and is neglected.
Approximating the readout as otherwise perfect and approximating the threshold of Supplementary Section~\ref{App:click_statistics} as a $\mathrm{B}$ assignment whenever $\ket{\mathrm{B}}$ is occupied for at least half of the window, the probability of missing an excursion to \(\ket{\mathrm{B}}\) is
\begin{equation}
p_{\mathrm{FN}}=1-\frac{1}{2}e^{-u/2}-\frac{e^{-u/2}-e^{-u}}{u},
\qquad u=T_{\mathrm{int}}/\tau_{\mathrm{B}}.
\label{eq:p_FN_from_Tint_over_tau_B}
\end{equation}
For \(T_{\mathrm{int}}=\qty{200}{\nano\second}\) and \(\tau_{\mathrm{B}}=\qty{5}{\micro\second}\), we obtain \(p_{\mathrm{FN}}\approx \qty{2.47}{\percent}\).
A longer integration time increases \(p_{\mathrm{FN}}\) and consequently shifts the predicted quantum-jump transition to higher \(\lambda\). For \(T_{\mathrm{int}}=\qty{320}{\nano\second}\), used in the real-time tomography of Fig.~\fullref{fig:first_transition}{a,b}, we obtain \(p_{\mathrm{FN}}\approx\qty{3.92}{\percent}\) and \(\reallambda{1}\approx1.32\), compared with \(\reallambdaval{1}\) at \(\qty{200}{\nano\second}\). To maintain consistency with the experimental conditions under which all transition locations were extracted, we use \(T_{\mathrm{int}}=\qty{200}{\nano\second}\) for all realistic-model predictions.

\paragraph{Ensemble dynamics.}
The unconditional evolution is unaffected by detector inefficiency: the true measurement backaction occurs regardless of whether the detector correctly reports it. The ensemble Liouvillian remains equation~\eqref{eq:full_lindblad}.

\paragraph{Postselected dynamics.}
For evolution conditioned on no clicks, detector errors modify the dynamics in two ways.

First, consider false positives. A postselected trajectory ends at a click, which can be due either to an excursion or to a false positive. In the ideal case, the no-click survival probability decreases at a rate $\alpha\langle 0|\hat{\rho}|0\rangle$ due to clicks, captured by the anticommutator $-\frac{\alpha}{2}\{|0\rangle\langle 0|, \hat{\rho}\}$. False positives provide an additional, state-independent channel for trajectory termination, contributing a uniform decay $-\kappa_\mathrm{FP}\hat{\rho}$.

Second, consider false negatives. When an excursion produces no click, the system experiences the measurement backaction---projection onto $|0\rangle$---yet the trajectory continues. In the ideal postselected dynamics, the reset term $\alpha\,|0\rangle\langle 0|\hat{\rho}|0\rangle\langle 0|$ is absent because every excursion produces a click that terminates the trajectory. With false negatives, a fraction $p_\mathrm{FN}$ of excursions produce no click, so this term is partially restored: $p_\mathrm{FN}\,\alpha\,|0\rangle\langle 0|\hat{\rho}|0\rangle\langle 0|$.

Combining these effects, the postselected Liouvillian becomes
\begin{equation}
\label{eq:postselected_lindblad_with_inefficiency}
\widetilde{\mathcal{L}}_\mathrm{P}[\hat{\rho}] = -\frac{i}{\hbar}[\hat{H}_\mathrm{d},\hat{\rho}] - \frac{\alpha}{2}\{\ketbra{0}{0}, \hat{\rho}\} - \kappa_\mathrm{FP}\hat{\rho} + p_\mathrm{FN}\,\alpha\,\ketbra{0}{0}\hat{\rho}\ketbra{0}{0} + \frac{1}{2T^\phi_\mathrm{S}}\mathcal{D}[\hat{\sigma}_\mathrm{z}]\hat{\rho} + \frac{1}{T_\mathrm{S}^1}\mathcal{D}[\ketbra{0}{1}]\hat{\rho}.
\end{equation}
The eigenvalue structure of $\widetilde{\mathcal{L}}_\mathrm{P}$ determines the quantum-jump and state-freezing transitions, as analysed in Section~\ref{App:transition_locations}.

\section{The Transition Locations in the Realistic Model}
\label{App:transition_locations}

We analyse how decoherence modifies the locations of the dynamical transitions, building on the realistic model of Section~\ref{App:realistic_model}. Since the eigenvalue structure of the relevant Liouvillians fully determines the transitions, we compute the critical measurement strengths numerically by exact diagonalisation.

\subsection{Postselected Liouvillian with Decoherence}
\label{App:transition_locations Postselected Liouvillian with Decoherence}

The quantum-jump and state-freezing transitions are determined by the no-click dynamics, governed by the postselected Liouvillian $\widetilde{\mathcal{L}}_\mathrm{P}$ of equation~\eqref{eq:postselected_lindblad_with_inefficiency}. To analyse its eigenvalue structure, we parametrise the density matrix in the $\ket{1}$--$\ket{0}$ basis as
\begin{equation}
\hat{\rho}_\mathrm{S}(t)=\frac{1}{2}\begin{pmatrix}
P^{(0)} + z & x \\
x & P^{(0)} - z
\end{pmatrix},
\label{eq:density-matrix-parametrised}
\end{equation}
where $P^{(0)} = \mathrm{Tr}[\hat{\rho}_\mathrm{S}]$ is the no-click survival probability, $z = \mathrm{Tr}[\hat{\rho}_\mathrm{S} \hat{\sigma}_\mathrm{z}]$, and $x = \mathrm{Tr}[\hat{\rho}_\mathrm{S} \hat{\sigma}_\mathrm{x}]$. The coherence $x$ is real since the dynamics preserve the $XZ$-plane of the Bloch sphere.

Converting equation~\eqref{eq:postselected_lindblad_with_inefficiency} to the vector representation $\boldsymbol{s} = [P^{(0)}, x, z]^T$ yields
\begin{equation}
\label{eq:s_vector_time_evolution}
\frac{d\boldsymbol{s}}{dt} = \widetilde{\mathcal{L}}_\mathrm{P} \, \boldsymbol{s},
\end{equation}
with matrix representation
\begin{equation}
\widetilde{\mathcal{L}}_\mathrm{P} = \begin{pmatrix}
-\frac{\alpha_\mathrm{eff}}{2} - \kappa_\mathrm{FP} & 0 & \frac{\alpha_\mathrm{eff}}{2}\\[4pt]
0 & -\gamma_2 - \frac{\alpha}{2} - \kappa_\mathrm{FP} & {-}\Omega_\mathrm{S} \\[4pt]
\frac{\alpha_\mathrm{eff}}{2} - \gamma_1 & {+}\Omega_\mathrm{S} & -\frac{\alpha_\mathrm{eff}}{2} - \gamma_1 - \kappa_\mathrm{FP}
\end{pmatrix}.
\label{eq:LP_matrix}
\end{equation}
Here, $\alpha_\mathrm{eff} = \alpha(1 - p_\mathrm{FN})$ is the effective measurement rate, $\gamma_1 = 1/T^1_\mathrm{S}$ is the relaxation rate, and $\gamma_2 = 1/T^\phi_\mathrm{S} + 1/(2T^1_\mathrm{S})$ is the total decoherence rate (including the contribution from relaxation).

The eigenvalues of $\widetilde{\mathcal{L}}_\mathrm{P}$ determine the transition locations. We order them by their real parts: $\operatorname{Re}(e_1) \leq \operatorname{Re}(e_2) \leq \operatorname{Re}(e_3) < 0$, where $e_3$ is always real while $e_1$ and $e_2$ form a complex-conjugate pair in the oscillatory regime where $\lambda < \reallambda{1}$.

\subsection{Splitting of the Quantum-Jump Transition in the Presence of Decoherence}
\label{App:first_transition_splitting}

The ideal model of ref.~\cite{snizhko_quantum_2020} (see also Fig.~\ref{fig:schematic_and_transitions} and Supplementary Section~\ref{App:ideal_model}) predicts that the transition into the continuous quantum-jump regime is marked by two concurrent features: the opening of a forbidden region and the cessation of oscillations in the postselected dynamics.
Our experimental data reveal that these two features decouple and occur at distinct measurement strengths, $\obslambdaeq{4}$ and $\obslambdaeq{1}$, respectively.
Here we discuss the origin of this splitting, show that the observed ordering is the only one possible, and give a quantitative criterion for the opening of the forbidden region.

In the ideal model for pure states, the dynamics are restricted to the unit circle in the $XZ$-plane of the Bloch sphere, cf.~Fig.~\ref{fig:schematic_and_transitions}.
The no-click evolution generated by equation~\eqref{Eq:non-hermitian hamiltonian} is a deterministic, continuous flow of the single coordinate $\theta$ on this circle, and such a flow has only two possibilities: without a fixed point it is periodic and traverses every angle, and with a fixed point it converges to it and never completes a revolution.
Oscillation and the absence of a forbidden region are therefore the same statement, and both end at the same $\lambda$, namely $\critlambda{1}=1$, where the fixed point $\theta_+$ appears (equation~\eqref{eq:theta_of_t_eq_of_motion}).

Decoherence removes this identity by making the conditioned state mixed: it is no longer confined to the unit circle and can lie inside the disc, so the flow acquires a second, radial coordinate (Fig.~\fullref{fig:second_transition}{a}).
The postselected dynamics of equation~\eqref{eq:LP_matrix} remain deterministic and continuous, but the trajectory now approaches an attracting fixed point $\mathbf{R}_\infty = (X_\infty, Z_\infty)$ inside the disc, given by the eigenvector of $\widetilde{\mathcal{L}}_\mathrm{P}$ belonging to the slowest eigenvalue $e_3$ (Supplementary Section~\ref{App:transition_locations Postselected Liouvillian with Decoherence}).
The approach is a spiral while $e_1,e_2$ are complex, i.e.\ for $\lambda<\reallambda{1}$, and non-oscillating for $\lambda>\reallambda{1}$.
Whether such a trajectory explores all the polar angles $\theta$ or is characterised by a forbidden region is now a separate question: does the spiral encircle the centre of the disc, or not?
The ordering of the two features is nevertheless fixed. Once oscillations have ceased, the trajectory converges to the fixed point without winding and the forbidden region is certainly open, so that
\begin{equation}
\label{eq:fr_ordering}
\generallambda{4} \le \generallambda{1},
\end{equation}
with equality in the ideal model and strict inequality as soon as the state can leave the circle.

The location of $\generallambda{4}$ follows from the asymptotic form of the spiral (Supplementary Section~\ref{App:transition_locations_The_Second_Transition}): the distance to the fixed point contracts as $A(t)\simeq A_0\,e^{-(e_3-\operatorname{Re}(e_2))t}$, where $A_0$ is the distance of the initial Bloch vector from the fixed point, while the angle about it advances at the rate $\operatorname{Im}(e_2)$.
The spiral encircles the centre of the disc only when its amplitude exceeds the distance $\|\mathbf{R}_\infty\|=\sqrt{X_\infty^2+Z_\infty^2}$ of the fixed point from the centre, so the number of complete windings it performs is approximately
\begin{equation}
\label{eq:fr_winding_estimate}
N \simeq \frac{\operatorname{Im}(e_2)}{2\pi}\;\frac{\ln\!\left(A_0/\|\mathbf{R}_\infty\|\right)}{e_3-\operatorname{Re}(e_2)}.
\end{equation}
The forbidden region opens where $N$ falls below unity.
Pure dephasing raises the contraction rate of the spiral $e_3-\operatorname{Re}(e_2)$, which reduces $N$ and drives the forbidden-region threshold to lower $\lambda$. $A_0$ depends on the initial state, so $\generallambda{4}$ is not fully determined by the spectrum, in contrast to the other transitions.
For the experimental parameters, equation~\eqref{eq:fr_winding_estimate} gives $N=1$ at $\lambda=0.53$ for preparation in $\ket{1}$ and at $0.56$ for $\ket{0}$.
The exact thresholds, obtained by integrating equation~\eqref{eq:LP_matrix} and locating the $\lambda$ at which the total angle swept by the trajectory drops below $2\pi$, are $\reallambda{4}=\reallambdaval{4}$ and $0.51$, respectively (Fig.~\ref{fig:transition_locations}).
Both lie far below $\reallambda{1}=\reallambdaval{1}$: between the two thresholds the trajectory still spirals, but no longer around the centre, which is the regime seen at $\lambda=0.53$ and $0.91$ in Fig.~\fullref{fig:second_transition}{a}.

The splitting is thus a direct consequence of the postselected dynamics generating mixed states.
A natural question is which of the two features should define the transition into the continuous quantum-jump regime.
We identify it with the cessation of oscillations, for two reasons.
First, the opening of the forbidden region is a property of a specific trajectory: it depends on the initial state, whereas the cessation of oscillations is a property of the spectrum of $\widetilde{\mathcal{L}}_\mathrm{P}$ and is independent of the initial state.
Second, the notion of a continuous quantum jump implies a non-oscillatory approach to the fixed point, in the sense of the continuous quantum jumps observed in ref.~\cite{minev_catch_2019}.
These considerations make $\generallambda{1}$ the natural threshold to associate with the onset of the quantum-jump regime, and $\generallambda{4}$ a distinct, geometric threshold that is decoupled from it by decoherence.

\subsection{The Quantum-Jump Transition: Cessation of Oscillations}
\label{App:transition_locations_The_First_Transition}
\begin{figure}[h!]
    \centering
    \includegraphics[width=0.9\linewidth]{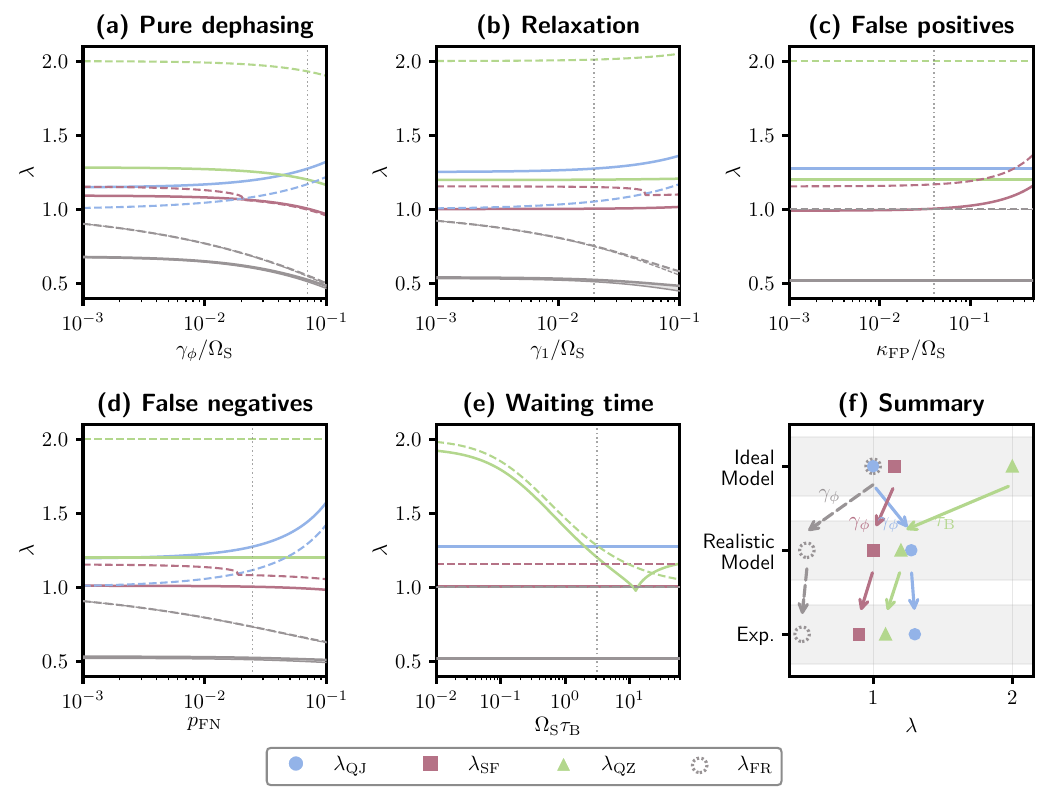}
    \caption{
    \textbf{Dependence of transition locations on decoherence parameters.}
    Dashed lines show numerical results from the realistic model with the indicated parameter varied and all other independent error parameters set to zero; solid lines show results with all other parameters held at experimental values.
The quantum-jump, state-freezing and quantum Zeno transitions are shown in blue, red and green, respectively. Grey curves show the opening of the forbidden region $\generallambda{4}$, a geometric feature of the no-click trajectory rather than a property of the spectrum. It therefore depends on the initial state.
The grey band spans preparation in $\ket{1}$ (as in Fig.~\fullref{fig:second_transition}{a}, upper edge), and in $\ket{0}$ (lower edge).
Vertical dotted lines indicate experimental parameter values.
\textbf{(a)} Effect of pure dephasing rate $\gamma_\phi/\Omega_\mathrm{S}$, with $\gamma_\phi=1/T^\phi_\mathrm{S}$.
\textbf{(b)} Effect of relaxation rate $\gamma_1/\Omega_\mathrm{S}$.
\textbf{(c)} Effect of false positive rate $\kappa_\mathrm{FP}/\Omega_\mathrm{S}$.
\textbf{(d)} Effect of false negative probability $p_\mathrm{FN}$.
\textbf{(e)} Effect of dimensionless waiting time $\Omega_\mathrm{S}\tau_\mathrm{B}$.
Although $p_\mathrm{FN}$ depends on $\tau_\mathrm{B}$ (see equation~\eqref{eq:p_FN_from_Tint_over_tau_B}), we hold it fixed in this panel to isolate the effect of $\tau_\mathrm{B}$.
\textbf{(f)} Summary comparing ideal values ($\critlambda{1}=1,\critlambda{2}=2/\sqrt{3},\critlambda{3}=2$), realistic model predictions ($\reallambda{1}=\reallambdaval{1},\reallambda{2}=\reallambdaval{2},\reallambda{3}=\reallambdaval{3}, \reallambda{4}=\reallambdaval{4}$), and experimental values ($\obslambdafull{1}, \obslambdafull{2}, \obslambdafull{3}, \obslambdafull{4}$).
Each arrow linking the ideal to the realistic model is labelled with the parameter that dominates that shift.
The uncertainties on the observed values are smaller than the marker size.
The realistic model reproduces the experimentally observed reordering of the transitions, with state freezing preceding the onset of quantum Zeno dynamics, followed by the quantum-jump transition.
This reordering reflects the distinct sensitivities of the transitions to different imperfections.
Detector errors leave the quantum Zeno transition unchanged \textbf{(c,d)}, since the physical backaction occurs whether or not the detector reports it correctly. False positives shift every eigenvalue of the postselected Liouvillian by the same amount, so they leave $\generallambda{1}$ and $\generallambda{4}$ unchanged as well, and shift only $\generallambda{2}$ \textbf{(c)}.
Conversely, the waiting time $\tau_\mathrm{B}$ shifts only the quantum Zeno transition \textbf{(e)}: it enters the quantum-jump, state-freezing and forbidden-region thresholds solely through $p_\mathrm{FN}$ (equation~\eqref{eq:p_FN_from_Tint_over_tau_B}).
}
    \label{fig:transition_locations}
\end{figure}

The quantum-jump transition marks the boundary between oscillatory and non-oscillatory no-click dynamics. It occurs when $e_1$ and $e_2$ transition from a complex-conjugate pair to distinct real values, passing through an exceptional point of $\widetilde{\mathcal{L}}_\mathrm{P}$. This requires the characteristic polynomial $\mathcal{C}_\mathrm{P}(\Lambda) = \det(\widetilde{\mathcal{L}}_\mathrm{P} - \Lambda\mathbb{I})$ to have a repeated root, i.e., $\mathcal{C}_\mathrm{P}(\Lambda) = 0$ and $\mathcal{C}_\mathrm{P}'(\Lambda) = 0$ simultaneously. The resulting $\reallambda{1}$ rises with finite decoherence, with pure dephasing providing the dominant contribution, as shown in Fig.~\ref{fig:transition_locations}.
False positive events appear as a state-independent term in $\widetilde{\mathcal{L}}_\mathrm{P}$, corresponding to a diagonal shift in the matrix representation of equation~\eqref{eq:LP_matrix}, which leaves the exceptional point determining $\reallambda{1}$ unaffected. False negatives, in contrast, allow excursions to go unobserved, sustaining oscillations up to higher measurement strengths and thereby raising $\reallambda{1}$.
The physical origin of the pure-dephasing-induced shift of $\reallambda{1}$ can be viewed as an additional unobserved monitoring channel, not excluded by postselection. Using $\mathcal{D}[\hat{\sigma}_\mathrm{z}]=4\mathcal{D}[\ketbra{0}{0}]$, equation~\eqref{eq:postselected_lindblad_with_inefficiency} becomes
\begin{equation}
\label{eq:rearranged_postselected_lindblad_with_dephasing}
\frac{d\hat{\rho}}{dt}= \widetilde{\mathcal{L}}_\mathrm{P}[\hat{\rho}]
= -\frac{i}{\hbar}[\hat{H}_\mathrm{d},\hat{\rho}]
+ \left(\frac{2}{T^\phi_\mathrm{S}} + p_\mathrm{FN}\alpha\right)\ketbra{0}{0}\hat{\rho}\ketbra{0}{0}
- \frac{1}{2}\left(\alpha+\frac{2}{T^\phi_\mathrm{S}}\right)\{\ketbra{0}{0},\hat{\rho}\}
- \kappa_\mathrm{FP}\hat{\rho}
+ \frac{1}{T^1_\mathrm{S}}\mathcal{D}[\ketbra{0}{1}]\hat{\rho}.
\end{equation}
This form exposes two competing effects of pure dephasing. The term $- \frac{1}{2}\left(\alpha+\frac{2}{T^\phi_\mathrm{S}}\right)\{\ketbra{0}{0},\hat{\rho}\}$ strengthens the no-click backaction. Taken alone, it would lower $\reallambda{1}$. At the same time, the term $\frac{2}{T^\phi_\mathrm{S}}\ketbra{0}{0}\hat{\rho}\ketbra{0}{0}$ adds an unobserved measurement channel. Defining $q \equiv \frac{2/T^\phi_\mathrm{S} +p_\mathrm{FN}\alpha}
{\alpha+2/T^\phi_\mathrm{S}}$, equation~\eqref{eq:rearranged_postselected_lindblad_with_dephasing} interpolates between the ideal no-click generator at $q=0$ and the full
ensemble Liouvillian at $q=1$, as in the hybrid-Liouvillian formalism of
ref.~\cite{minganti_hybrid-liouvillian_2020}. The unobserved measurement contribution shifts the oscillatory exceptional point towards that of the full Liouvillian and therefore competes with the direct increase of the no-click backaction.
A perturbative expansion of $\reallambda{1}$ about the ideal critical point in the presence of weak pure dephasing with all other imperfections set to zero resolves this competition, showing that the net correction is positive:
\begin{equation}
\label{eq:first_transition_perturbative_shift}
\reallambda{1} =
1+\left(\frac{27}{32}\right)^{1/3}\epsilon^{\frac{2}{3}} + \cdots,
\qquad
\epsilon = \frac{1}{\Omega_\mathrm{S} T^\phi_\mathrm{S}}.
\end{equation}
Thus, weak pure dephasing raises the quantum-jump transition location.
Interestingly, a finite dephasing rate also reduces the order of the exceptional point (the number of eigenvalues coalescing at the exceptional point) from three to two: only the complex-conjugate pair coalesces, while the longest-lived real eigenvalue remains isolated, giving rise to an attracting fixed point inside the Bloch sphere, consistent with the discussion in ref.~\cite{pavlov_topological_2025}. 
For our experimental parameters ($T^1_\mathrm{S} = \qty{81}{\micro\second}$, $T^\phi_\mathrm{S} = \qty{23}{\micro\second}$, $\Omega_\mathrm{S}/2\pi =  \qty{100}{\kilo\Hz}$, $p_{\mathrm{FN}} = \qty{2.47}{\percent}$), numerical diagonalisation yields $\reallambda{1} = \reallambdaval{1}$.

\begin{figure}[t]
  \centering
  \includegraphics[width=\linewidth]{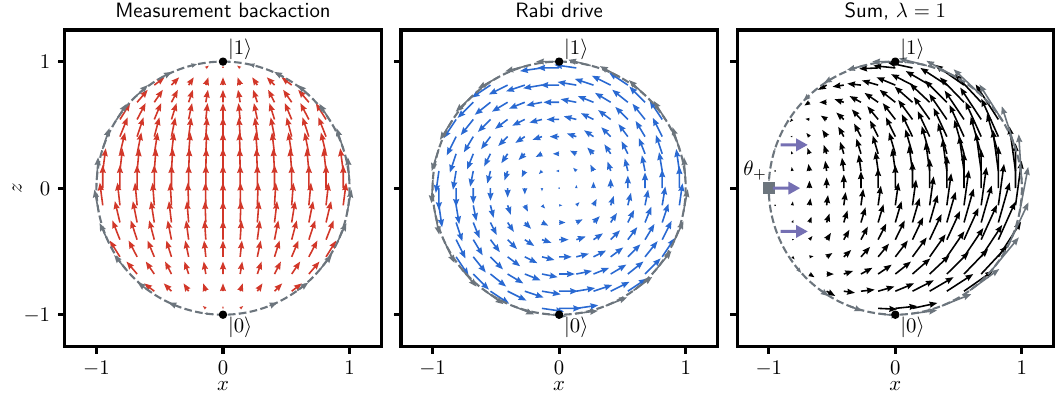}
  \caption{
  Velocity of the conditional Bloch vector in the $XZ$-plane for pure states (grey, on the circle) and mixed states (coloured, interior).
  \emph{Left:} no-click measurement backaction, whose component towards \(\ket{1}\) near the equator remains unchanged as the Bloch-vector length decreases; inside the Bloch sphere it also acquires a repurifying component.
  \emph{Centre:} Rabi drive, a rigid rotation whose speed scales with the Bloch vector length.
  \emph{Right:} their sum at $\lambda=1$, with the fixed point $\theta_+$ on the
  equator; purple arrows indicate the action of dephasing, which
  displaces the state from the surface into the region where the flow still
  circulates and hence sustains the oscillatory regime to higher $\lambda$ in the realistic model.
  Arrow lengths and directions are qualitative and chosen to illustrate the flow mechanism rather than quantitative values.
  }
  \label{fig:backaction_rabi_fields_on_bloch_circle}
\end{figure}

For a physical picture, consider the dynamics near the ideal fixed point at $\lambda = 1$, where $\theta_+ = -\pi/2$. For a pure state on the Bloch-sphere
surface, the Rabi drive and the no-click backaction cancel exactly at this
point. Dephasing shortens the Bloch vector and displaces the state into the
interior of the sphere. This breaks the cancellation because the two
contributions scale differently with the Bloch-vector length~$\|\mathbf{R}\|$: the Rabi
drive generates a rotation with velocity proportional to~$\|\mathbf{R}\|$
(Fig.~\ref{fig:backaction_rabi_fields_on_bloch_circle}, centre), whereas near the equator the no-click backaction has a
component towards $|1\rangle$ that is independent of~$\|\mathbf{R}\|$ (Fig.~\ref{fig:backaction_rabi_fields_on_bloch_circle}, left).
Thus, once $\|\mathbf{R}\| < 1$, the no-click backaction initially dominates, pushing the
state towards $\ket{1}$ and repurifying it. This restores the Rabi
contribution and produces a damped spiral with finite oscillation frequency at
$\lambda = 1$, where the ideal dynamics have already reached their exceptional
point. Dephasing therefore allows the oscillatory regime to persist to larger
$\lambda$, shifting the quantum-jump transition upwards. An additional intuitive angle can be gained from a phase-flip unravelling of the dephasing channel, where unobserved phase flips can enable trajectories to complete a cycle around the Bloch sphere without a click being recorded, thus sustaining oscillations at higher~$\lambda$.

\subsection{Defining the State-Freezing Transition in the Presence of Decoherence}
\label{App:tau_subtleties}

The state-freezing transition is identified via a divergence in the dwell-time density $\tau_\theta(\theta)$, defined through equations~\eqref{eq:tau_theta_definition}--\eqref{eq:tau_theta_P0_theta_over_theta_dot} for the ideal model. In the case of the realistic model that features decoherence, $\theta$ is not the natural parameter for the system's no-click trajectory since multiple points on the trajectory may correspond to the same value of $\theta$ (cf.~Fig.~\fullref{fig:second_transition}{a}).
The natural trajectory parameter is the arc length $l(t)$ spanned by the dynamics defined through
\begin{equation}
\label{eq:arc_length_velocity}
    \dot{l}(t) = \sqrt{\dot{X}^2(t)+\dot{Z}^2(t)},
\end{equation}
where $X(t) = \mathrm{Tr}[\hat{\rho}_\mathrm{S} \hat{\sigma}_\mathrm{x}] / P^{(0)} = x / P^{(0)}$ and $Z(t) = \mathrm{Tr}[\hat{\rho}_\mathrm{S} \hat{\sigma}_\mathrm{z}] / P^{(0)} = z / P^{(0)}$ are the qubit state coordinates in the $XZ$-plane of the Bloch sphere, rescaled by the no-click survival probability $P^{(0)}(t) = \mathrm{Tr}[\hat{\rho}_\mathrm{S}]$ (see equation~\eqref{eq:density-matrix-parametrised}).
Then one can define the dwell-time density in the trajectory interval $\Delta l \rightarrow 0$
\begin{equation}
\label{eq:tau_s_definition}
    \tau_l(l) = \frac{P^{(0)}(l)}{\dot{l}},
\end{equation}
where $P^{(0)}(l) = P^{(0)}(t(l))$ is the probability of a no-click trajectory reaching the arc length $l$.

With the fundamental quantity being the probability of experiencing no clicks up to time $t$, $P^{(0)}(t)$, one can relate $\tau_\theta(\theta)$ (dwell time per $\Delta \theta$) and $\tau_l(l)$ (dwell time per $\Delta l$) via
\begin{equation}
\label{eq:tau_theta_via_tau_s}
    \tau_\theta(\theta) = \frac{P^{(0)}(\theta)}{|\dot{\theta}(t)|} = \frac{P^{(0)}(l)}{\dot{l}(t)} \frac{1}{|d\theta/dl|} = \frac{\tau_l(l)}{|d\theta/dl|}.
\end{equation}
It is important to mark here the change of variable  $\tau_\theta(\theta(l)) = \tau_l(l)/|d\theta/dl|$ and the difference in physical meaning between $\tau_\theta(\theta)$ (dwell time per $\Delta \theta$) and $\tau_l(l)$ (dwell time per $\Delta l$). It is the latter that leads to the appearance of the factor $d\theta/dl$.

Two important aspects are worth noticing when $\theta(l)$ dependence is non-monotonic (as is the case for spiralling trajectories in Fig.~\fullref{fig:second_transition}{a}).
First, equation~\eqref{eq:tau_theta_via_tau_s} should include a sum over all the points $l(t)$ corresponding to a given value of $\theta(t)$; i.e., a sum over the number of loops of the spiral. Second, $d\theta/dl = 0$ at the points where the spiral folds, creating divergences in $\tau_\theta(\theta)$. Evidently, these divergences are not associated with measurement-induced changes of dynamics, but are simply the consequences of parametrising the system trajectory via an unsuitable parameter $\theta$. While the first subtlety is trivial to account for, the second one warrants a brief discussion.

In the vicinity of a spiral fold, $\theta(l) - \theta(l_\mathrm{fold}) \sim (l - l_\mathrm{fold})^2$, so that $d\theta(l)/dl \sim (l - l_\mathrm{fold})$, leading to $\tau_\theta$ divergence at the fold:
\begin{equation}
    \tau_\theta(\theta)
    \sim \frac{\tau_l(l_\mathrm{fold})}{(l - l_\mathrm{fold})}
    \sim \frac{\tau_l(l_\mathrm{fold})}{\sqrt{|\theta - \theta_\mathrm{fold}|}}.
\end{equation}
When binning the data, $\tau_\theta^{\text{(binned)}} = \int_{\theta_1}^{\theta_2} d\theta \tau_\theta(\theta) / |\theta_2 - \theta_1| \sim 1/\sqrt{|\theta_2 - \theta_1|}$, smoothing the divergence due to the finite bin size. We have verified in our numerical simulations that for smaller bin size, the fold divergences do become visible. However, the bin size for the experimental data analysis is dictated by the single-measurement integration time $T_\mathrm{int}$ and cannot be taken arbitrarily small.

It is interesting to note that in contrast to the fold divergences, the asymptotic divergence close to the end of the no-click trajectory, $\tau_\theta(\theta) \sim |\theta - \theta_+|^{\xi}$ clearly manifests in the experimental data (Fig.~\fullref{fig:second_transition}{a}) despite the measured value of $\xi > -1/2$ (in principle, $\xi \to -1$ for $\lambda \to \infty$, but we do not approach such values in our experiments).
While mathematically this divergence is weaker than the fold divergences, physically the coefficient multiplying the diverging function happens to be much larger for the end-of-trajectory divergence, making it clearly pronounced in the binned data.

\subsection{The Location of the State-Freezing Transition in the Realistic Model}
\label{App:transition_locations_The_Second_Transition}

The state-freezing transition marks the onset of freezing dynamics near the fixed point $\theta_+$. The transition arises from the competition between two rates: the click rate at the fixed point $-e_3$ and the approach rate to the fixed point $-\operatorname{Re}(e_2) + e_3$. As a result of this competition, the dwell time near the fixed point scales as $\tau_l(l) \sim |l - l_\infty|^{\xi}$, where the critical exponent is
\begin{equation}
\xi = \frac{e_3}{\operatorname{Re}(e_2) - e_3} - 1 = \frac{2e_3 - \operatorname{Re}(e_2)}{\operatorname{Re}(e_2) - e_3}.
\label{eq:ksi_from_eigenvalues}
\end{equation}
The transition occurs when $\xi = 0$, i.e., when $\operatorname{Re}(e_2) = 2e_3$.

This relation is obtained as follows. The trajectory parameters $\boldsymbol{s}(t) = [P^{(0)}(t), x(t), z(t)]^T$ can be expressed as
\begin{equation}
    \boldsymbol{s}(t) = \sum_{k=1}^3 \boldsymbol{s}_k e^{e_k t},
\end{equation}
where $\boldsymbol{s}_k$ and $e_k$ are the eigenvectors and the eigenvalues of the Liouvillian~\eqref{eq:LP_matrix}.
Considering the limit $t\to \infty$ and using the hierarchy of the eigenvalues defined in Section~\ref{App:transition_locations Postselected Liouvillian with Decoherence}, one finds that 
\begin{align}
    P^{(0)}(t) & \approx C_3 e^{e_3 t},  \\
    X(t) = x/P^{(0)}(t) & \approx X_\infty + \Delta X(t) e^{(\operatorname{Re}(e_2)-e_3)t}, \\
    Z(t) = z/P^{(0)}(t) & \approx Z_\infty + \Delta Z(t) e^{(\operatorname{Re}(e_2)-e_3)t},
\end{align}
where $C_3, X_\infty$, and $Z_\infty$ are constants, while $\Delta X(t)$ and $\Delta Z(t)$ may be constants or periodic functions oscillating at a frequency $\operatorname{Im}(e_2)$. Consequently, the arc-length velocity~\eqref{eq:arc_length_velocity}
has an asymptotic envelope that scales as
\begin{equation}
\dot{l}(t) = \sqrt{\dot{X}(t)^2+\dot{Z}(t)^2}
\sim e^{[\operatorname{Re}(e_2)-e_3]t}.
\end{equation}
Using these relations and the definition of $\tau_l(l)$ (equation~\eqref{eq:tau_s_definition}), we obtain
\begin{equation}
\tau_l(l(t))
=\frac{P^{(0)}(t)}{\dot l(t)}
\sim e^{[2e_3-\operatorname{Re}(e_2)]t}.
\end{equation}
The remaining arc length to the fixed point satisfies
\begin{equation}
|l(t)-l_\infty|
=\int_t^\infty \dot l(t')\,dt'
\sim e^{[\operatorname{Re}(e_2)-e_3]t}.
\end{equation}
Combining these expressions gives the power-law envelope of the dwell-time density
\begin{equation}
\tau_l(l(t\to\infty))
\sim |l(t)-l_\infty|^\xi,
\qquad
\xi=\frac{2e_3-\operatorname{Re}(e_2)}
{\operatorname{Re}(e_2)-e_3},
\end{equation}
which in turn translates into $\tau_\theta(\theta) \sim |\theta - \theta_+|^{\xi}$ through equation~\eqref{eq:tau_theta_via_tau_s}.

In the ideal model, all eigenvalues are real for $\lambda > 1$:
\begin{equation}
e_3 = -\Omega_\mathrm{S}(\lambda - \sqrt{\lambda^2-1}), \quad
e_2 = -\Omega_\mathrm{S}\lambda, \quad
e_1 = -\Omega_\mathrm{S}(\lambda + \sqrt{\lambda^2-1}),
\end{equation}
and the condition $e_2 = 2e_3$ yields $\critlambda{2} = 2/\sqrt{3} \approx 1.15$. With finite decoherence, however, the transition location shifts (Fig.~\ref{fig:transition_locations}).
Notably, dephasing \emph{lowers} the location of the state-freezing transition, in contrast to its effect on the quantum-jump transition. 
False positives terminate no-click trajectories
without changing their evolution, increasing
$\gamma_0$ while leaving $\gamma_\mathrm{a}$ unchanged, shifting $\reallambda{2}$ upwards.

For our experimental parameters, numerical diagonalisation yields $\reallambda{2} = \reallambdaval{2}$, confirming that dephasing can reorder the transitions.

For weak pure dephasing, this shift can also be obtained
perturbatively. Setting the other imperfections to zero and expanding
the condition $\operatorname{Re}(e_2)=2e_3$ around the ideal value
$\critlambda{2}=2/\sqrt{3}$ gives
\begin{equation}
\label{eq:second_transition_perturbative_shift}
\reallambda{2}
=
\frac{2}{\sqrt{3}}
-\frac{7}{3}\epsilon
+\mathcal{O}(\epsilon^2),
\qquad
\epsilon=\frac{1}{\Omega_\mathrm{S}T^\phi_\mathrm{S}} .
\end{equation}
Thus weak pure dephasing shifts the state-freezing transition
downwards, opposite to its effect on the quantum-jump transition, cf. equation~\eqref{eq:first_transition_perturbative_shift}.

It is interesting to note that unlike the quantum-jump and quantum Zeno transitions, which are exceptional points where two eigenvalues coalesce and the evolution generator becomes non-diagonalisable, the $\xi = 0$ condition is not a degeneracy or exceptional point.
Since $e_3<0$, $\operatorname{Re}(e_2)=2e_3$ excludes $e_2=e_3$: at $\reallambda{2}$ the eigenvalues remain distinct and $\widetilde{\mathcal{L}}_\mathrm{P}$ remains diagonalisable with a nondegenerate spectrum.
Indeed, for our experimental parameters $\reallambda{2}<\reallambda{1}$, so $e_1,e_2$ are still a complex-conjugate pair when the state-freezing transition occurs. What becomes degenerate is a pair of physical rates --- the asymptotic click rate $-e_3$ and the approach rate $e_3-\operatorname{Re}(e_2)$ --- not the generator.

The state-freezing transition is, therefore, not an exceptional point of $\hat{H}_\mathrm{eff}$, $\widetilde{\mathcal{L}}_\mathrm{P}$ or $\mathcal{L}$. The spectrum is analytic through $\reallambda{2}$, and the transition is sharp in the observable rather than in the spectrum.

\subsection{The Location of the Quantum Zeno Transition in the Realistic Model}
\label{App:transition_locations_The_Third_Transition}

The quantum Zeno transition occurs in the ensemble dynamics, marking the boundary between oscillatory and overdamped relaxation to the steady state. Unlike the other two transitions, it is primarily affected by the finite waiting time $\tau_\mathrm{B}$ in the detector's excited state.

The ensemble dynamics including the detector's excited state are given by equation~\eqref{eq:S_evolution}. Extending the parametrisation~\eqref{eq:density-matrix-parametrised} to include the detector population, we define $\boldsymbol{v} = [p_\mathrm{B}, p_\mathrm{S}, x, z]^T$, where $p_\mathrm{S} = \mathrm{Tr}[\hat{\rho}_\mathrm{S}]$ is the population of the qubit manifold $\{\ket{0},\ket{1}\}$, so that $p_\mathrm{S} + p_\mathrm{B} = 1$. The evolution becomes $d\boldsymbol{v}/dt = \mathcal{L}_\mathrm{B} \boldsymbol{v}$, with
\begin{equation}
\mathcal{L}_\mathrm{B} = \begin{pmatrix}
-\frac{1}{\tau_\mathrm{B}} & \frac{\alpha}{2} & 0 & -\frac{\alpha}{2} \\[4pt]
\frac{1}{\tau_\mathrm{B}} & -\frac{\alpha}{2} & 0 & \frac{\alpha}{2} \\[4pt]
0 & 0 & -\gamma_2 - \frac{\alpha}{2} & {-}\Omega_\mathrm{S} \\[4pt]
-\frac{1}{\tau_\mathrm{B}} & \frac{\alpha}{2} - \gamma_1 & {+}\Omega_\mathrm{S} & -\frac{\alpha}{2} - \gamma_1
\end{pmatrix}.
\label{eq:averaged-evolution-superoperator}
\end{equation}
Note that the ensemble dynamics are independent of detector inefficiency, since the physical measurement backaction occurs regardless of detection outcome.

The quantum Zeno transition occurs when the characteristic polynomial $\mathcal{C}_\mathrm{B}(\Lambda) = \det(\mathcal{L}_\mathrm{B} - \Lambda\mathbb{I})$ has a repeated root. The waiting time $\tau_\mathrm{B}$ interpolates between two limits: for $\tau_\mathrm{B} \to 0$, the system returns instantly from $\ket{\mathrm{B}}$ to $\ket{0}$, recovering the standard ensemble dynamics with $\critlambda{3} = 2$ in the ideal case.
For $\tau_\mathrm{B} \to \infty$, the system never returns, and the ensemble dynamics reduce to the postselected dynamics, giving $\reallambda{3} \to \reallambda{1}$.
As shown in Fig.~\fullref{fig:transition_locations}{(e)}, $\reallambda{3}$ drops significantly even for modest $\tau_\mathrm{B}$.

To separate the additional effect of dephasing from this
waiting-time-induced shift, we expand the same repeated-root condition
for weak pure dephasing. Setting the other imperfections to zero and
fixing the waiting time at its experimental value, we obtain
\begin{equation}
\label{eq:third_transition_perturbative_shift}
\reallambda{3}
=
1.28
-1.16\,\epsilon
+\mathcal{O}(\epsilon^2),
\qquad
\epsilon=\frac{1}{\Omega_\mathrm{S}T^\phi_\mathrm{S}} .
\end{equation}
Thus, weak pure dephasing shifts the quantum Zeno transition downwards,
as it does the state-freezing transition,
cf.~equation~\eqref{eq:second_transition_perturbative_shift}, and unlike
the quantum-jump transition, which shifts upwards,
cf.~equation~\eqref{eq:first_transition_perturbative_shift}.
Together with the finite-\(\tau_\mathrm{B}\) shift discussed above, and false negatives further raising the quantum-jump threshold, this
opposite response to dephasing explains why the quantum-jump and quantum
Zeno transitions exchange order in the realistic model.
For our experimental parameters, exact numerical diagonalisation yields $\reallambda{3} = \reallambdaval{3}$.
\subsection{Comparison with the Observed Values}
\label{App:model_vs_experiment}

The observed $\obslambda{2}$ and $\obslambda{3}$ lie about \qty{10}{\percent} below the realistic-model values, and $\obslambda{4}$ about \qty{6}{\percent} below, whereas $\obslambda{1}$ agrees within its uncertainty.
Two inputs of the model rest on idealised descriptions of the detection.
First, $\alpha$ is calibrated with a hidden Markov model that treats each integration window as an instantaneous measurement (Supplementary Section~\ref{App:alpha_extraction}); an underestimate of $\alpha$ by a given fraction lowers all four observed thresholds by that fraction.
Second, $p_\mathrm{FN}$ is estimated from equation~\eqref{eq:p_FN_from_Tint_over_tau_B} rather than measured; $\reallambda{1}$ rises by about $0.03$ per percentage point of $p_\mathrm{FN}$, $\reallambda{2}$ and $\reallambda{4}$ fall by less than $0.003$, and $\reallambda{3}$ is unaffected.
We have not established that these effects account for the offsets.
\section{Extraction of Observed Transition Values}
\label{App:extracting_critical_lambdas}
In this section, we detail the procedures used to extract the observed transition locations $\obslambda{1}$, $\obslambda{2}$ and $\obslambda{3}$ from the raw trajectories and derived observables, expanding on the overview in the Methods section.
We also describe how we locate $\obslambda{4}$, where the forbidden region opens. This is not a transition point but a geometric threshold of a particular trajectory (Supplementary Section~\ref{App:first_transition_splitting}).

\subsection{Quantum-Jump Transition --- $\obslambda{1}$}
\label{App:extracting_QJ_transition_lambda}

\begin{figure}[h!]
    \centering
    \includegraphics{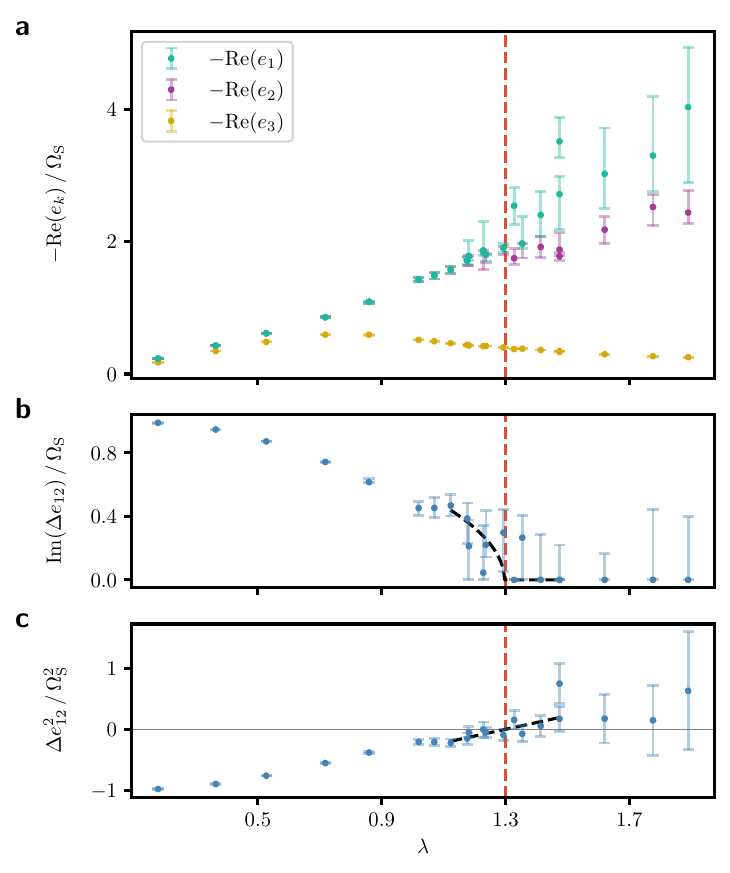}
    \caption{
    \textbf{Extracting $\obslambda{1}$ from the pole spectrum of the no-click survival probability $P^{(0)}(t)$.}
    \textbf{a,} Decay rates, given by the negative real part of the three poles, governing $P^{(0)}(t)$.
    \textbf{b,} Imaginary component of the pole splitting $\Delta e_{12} = (e_{2} - e_{1})/2$, corresponding to the frequency of the oscillatory component of $P^{(0)}(t)$.
    \textbf{c,} Squared pole splitting $\Delta e_{12}^2$.
    In all panels, the points are extracted from the experimental data using the matrix pencil method. The dashed black line in \textbf{c} shows a weighted linear fit of $\Delta e_{12}^2$ to equation~\eqref{eq:approx_transition_behaviour} near the critical point, while the dashed black curve in \textbf{b} shows the corresponding $\operatorname{Im}(\Delta e_{12})$ obtained from the same fit.
    The quantum-jump transition is identified by the sign change of $\Delta e_{12}^2$, with the zero crossing of the linear fit determining $\obslambdafull{1}$ (vertical red line).
    Error bars represent $\qty{68}{\percent}$ bootstrap confidence intervals, evaluated separately for each decay rate in \textbf{a}, propagated from the bootstrap distribution of $\Delta e_{12}^2$ in \textbf{b}, and symmetrised in \textbf{c}, using the same bootstrap samples throughout.
    }
    \label{fig:crit_lambda_1_with_poles}
\end{figure}

The matrix in equation~\eqref{eq:LP_matrix} has three eigenvalues and therefore predicts that, even in the presence of decoherence, the dynamics of $P^{(0)}(t)$ --- the probability that no detector clicks have occurred up to time $t$ --- are described by a linear combination of three complex exponentials. The quantum-jump transition is the point where two of these eigenvalues coalesce. In the Laplace domain, the eigenvalues correspond to poles; below the transition the pole set comprises a complex-conjugate pair $e_1, e_2$ and a real pole; at the transition the pair becomes real, leaving only real poles and hence purely decaying dynamics. The squared pole splitting $\Delta e_{12}^2 = \left[ (e_2 - e_1) / 2 \right] ^2$ is therefore negative below the transition and positive above it. Its zero crossing marks $\obslambda{1}$, as shown in Fig.~\fullref{fig:crit_lambda_1_with_poles}{c}.

We extract the three poles of $P^{(0)}(t)$ using the matrix pencil method~\cite{sarkar_using_1995}, a high-resolution technique for estimating the poles of complex exponentials, setting the model order to three.
Confidence intervals are estimated with a bootstrap procedure separately for each $\lambda$.
Two thousand bootstrap samples of $P^{(0)}(t)$ are generated by adding binomial shot noise, set by the experimental number of shots, to the measured data.
The matrix pencil method is then reapplied to every bootstrap sample independently. 
Consistent tracking of the pole labelling across different bootstrap iterations is enforced by a fixed sorting convention. $e_1$ and $e_2$ are ordered by real part; when they form a complex-conjugate pair, $e_2$ denotes the member with a positive imaginary part.
Since $P^{(0)}(t)$ is real, the matrix pencil method yields either real poles or complex-conjugate pairs. The squared splitting $\Delta e_{12}^2$ therefore has a well-defined sign in every bootstrap iteration.
The plotted points in Fig.~\ref{fig:crit_lambda_1_with_poles} are obtained by applying the matrix pencil method to the original data. Error bars are based on $68\%$ bootstrap confidence intervals, calculated separately for each decay rate in Fig.~\fullref{fig:crit_lambda_1_with_poles}{a} and for the imaginary pole splitting in Fig.~\fullref{fig:crit_lambda_1_with_poles}{b}. In Fig.~\fullref{fig:crit_lambda_1_with_poles}{c}, the same error bars are symmetrised and centred on the original-data estimates, providing uncertainties used to weight the linear fit that determines $\obslambda{1}$. The symmetrisation is justified by the approximately symmetric bootstrap distributions.
The extracted poles are sensitive to residual population in $\ket{\mathrm{B}}$ at the start of a run, which distorts the first points of $P^{(0)}(t)$. We therefore perform a real-time check before each run to verify that the system is not in $\ket{\mathrm{B}}$, and initiate the experiment only if this condition is satisfied. With this check, $\obslambda{1}$ remains robust when removing up to three time steps from either end of $P^{(0)}(t)$.
At the critical measurement strength $\obslambda{1}$, two eigenvalues coalesce, and the Liouvillian in equation~\eqref{eq:LP_matrix} acquires a $2\times2$ Jordan block structure. Expanding around this transition, the eigenvalue splitting depends on the perturbation $\delta = \lambda - \obslambda{1}$. While the individual eigenvalues also experience common-mode shifts, these cancel out when taking the difference between them. Consequently, the splitting is governed by the expansion $\Delta e_{12} = \sqrt{c_1 \delta + c_2 \delta^2 + \dots}$~\cite{welters_explicit_2011}, characteristic of second-order exceptional points. Its square therefore satisfies $\Delta e_{12}^2 = c_1\delta + O(\delta^2)$, justifying the linear approximation in equation~\eqref{eq:approx_transition_behaviour}.

Fitting the squared splitting $\Delta e_{12}^2$ to a linear equation avoids the numerical instability associated with the fitting of the square root dependence of the splitting $\Delta e_{12}$. The fitted zero crossing and its standard error give $\obslambda{1}$ and its uncertainty, respectively. Tests on simulated data confirm this procedure is unbiased.
The dashed black curve in Fig.~\fullref{fig:crit_lambda_1_with_poles}{b} is obtained from the square root of this linear fit. Since the expansion holds locally, the fit only uses a narrow window around the initial guess for $\obslambda{1}$. Varying the size of this window over a reasonable range leaves $\obslambda{1}$ unchanged within uncertainties, yielding $\obslambdafull{1}$.

\subsection{State-Freezing Transition --- $\obslambda{2}$}
\label{App:extracting_second_transition_lambda}

The state-freezing transition is identified from the zero crossing of the critical exponent $\xi$, which governs the dwell-time density near the fixed point, $\tau_l(l)\propto(l_\infty-l)^{\xi}$ (Supplementary Section~\ref{App:transition_locations_The_Second_Transition}).
The angular dwell-time histograms in Fig.~\fullref{fig:second_transition}{a}, constructed as described in the Methods section, visualise this behaviour; Fig.~\ref{fig:second_transition_bloch_grid} shows them at $15$ measured values of $\lambda$ from $0.18$ to $1.23$.
We extract $\xi$ directly from the two late-time rates that determine it,
\begin{equation}
\label{eq:xi_from_rates}
\xi=\frac{\gamma_0}{\gamma_\mathrm{a}}-1,
\end{equation}
where the approach rate $\gamma_\mathrm{a}=e_3-\operatorname{Re}(e_2)$ and the click rate at the fixed point $\gamma_0=-e_3$ reproduce equation~\eqref{eq:ksi_from_eigenvalues}.
We obtain $\gamma_0$ from the slowest pole $e_3$ of an order-3 matrix-pencil fit to $P^{(0)}(t)$ over the whole record.

We obtain $\gamma_\mathrm{a}$ from the conditional Bloch vector $\mathbf{R}(t)=(X(t),Z(t))$ reconstructed by tomography at each $\lambda$.
This vector is the ratio of the unnormalised components $(x(t),z(t))$ to $P^{(0)}(t)$, and each of these is a sum of the three modes $e^{e_k t}$ of the postselected Liouvillian (Supplementary Section~\ref{App:transition_locations_The_Second_Transition}).
We therefore fit a survival-weighted form, in which the ratio is removed.
The fixed point $\mathbf{R}_\infty$ is given by the normalised eigenvector of $e_3$.
We define $w(t)=P^{(0)}(t)\,e^{\gamma_0 t}/C_3$, where $C_3$ is the amplitude of the slowest mode in $P^{(0)}(t)$.
This mode then cancels in $w(t)\left[\mathbf{R}(t)-\mathbf{R}_\infty\right]$, leaving at all times only the exponents $e_1-e_3$ and $e_2-e_3$:
\begin{equation}
\label{eq:weighted_fit}
w(t)\left[\mathbf{R}(t)-\mathbf{R}_\infty\right]=e^{-\gamma t}\left[\mathbf{a}\cosh(\beta t)+\mathbf{b}\,\frac{\sinh(\beta t)}{\beta}\right],
\end{equation}
where $\beta$ is real for a monotonic approach and imaginary for a spiral, corresponding to real or complex-conjugate $e_1,e_2$.
We fit $\mathbf{R}_\infty$, $\mathbf{a}$, $\mathbf{b}$, $\gamma$ and $\beta$ at each $\lambda$, with $w(t)$ fixed by the measured $P^{(0)}(t)$ and its slowest pole.
Since $w(t)\to1$ at late times, the deviation decays at the slower of the two rates $\gamma\pm\operatorname{Re}(\beta)$, which gives the approach rate $\gamma_\mathrm{a}=\gamma-\operatorname{Re}(\beta)$.
The fitted curves are overlaid on the measured data in Fig.~\ref{fig:second_transition_bloch_grid}.

\begin{figure}
    \centering
    \includegraphics{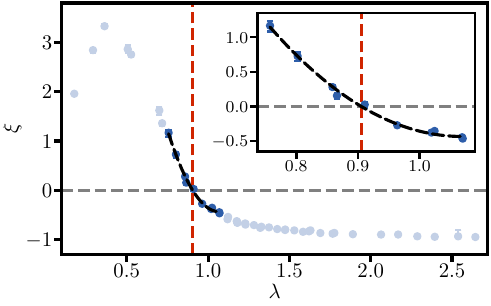}
    \caption{
    \textbf{Extraction of $\obslambda{2}$ from the critical exponent $\xi$.}
    Critical exponent $\xi=\gamma_0/\gamma_\mathrm{a}-1$ at all measured $\lambda$, with the ten scan points nearest the sign change in full colour and the others faded; the inset shows those ten points.
    Error bars are $68\%$ bootstrap intervals.
    The dashed black line shows the local quadratic fit to the ten points.
    The state-freezing transition is identified at $\obslambdafull{2}$ (dashed red vertical line), the zero crossing of the fit (horizontal grey dashed line).
    }
    \label{fig:second_transition_xi_fit_full}
\end{figure}

\begin{figure}
    \centering
    \includegraphics[width=\textwidth]{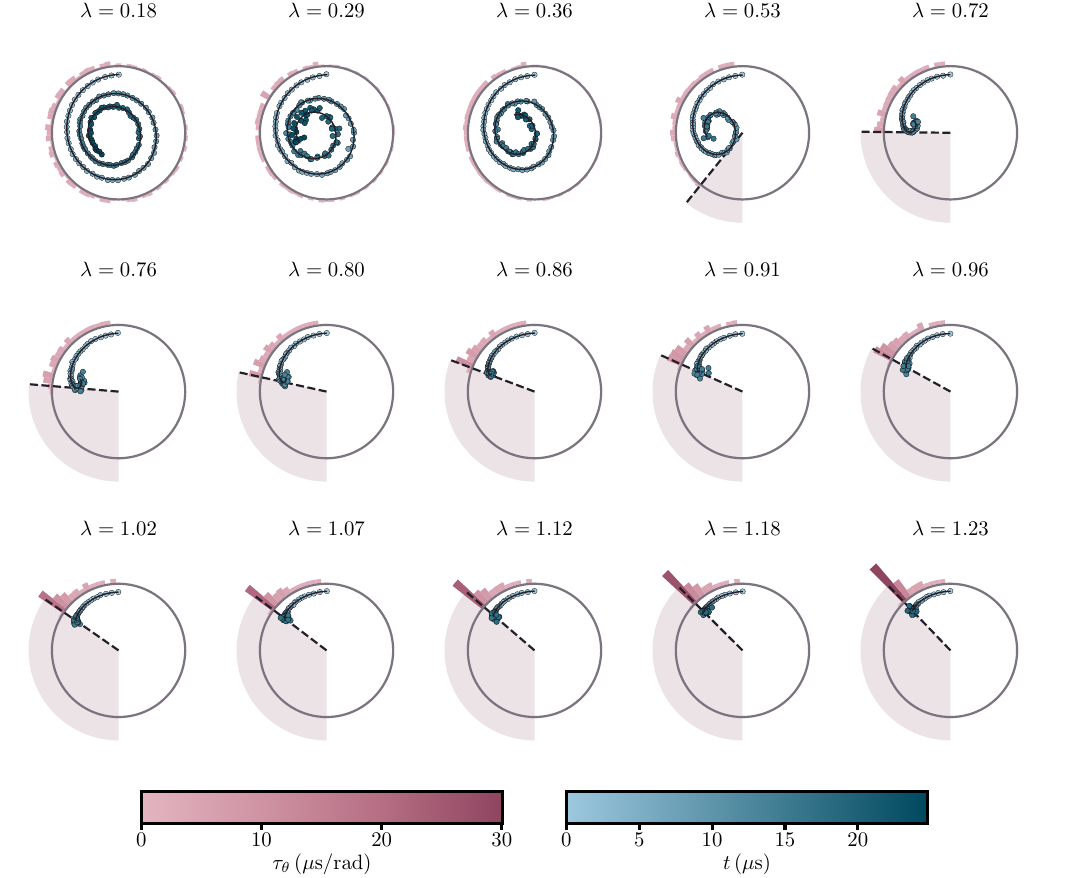}
    \caption{
    \textbf{No-click trajectories and angular dwell time across the opening of the forbidden region and the state-freezing transition.}
    Bloch cross-sections as in Fig.~\fullref{fig:second_transition}{a}, at $15$ measured values of $\lambda$ from $0.18$ to $1.23$.
    Circles: reconstructed no-click trajectory $\mathbf{R}(t)$ with $P^{(0)}(t)\geq0.01$, the readouts used in the fit, coloured by time.
    Solid black line: survival-weighted spiral fit (equation~\eqref{eq:weighted_fit}).
    The same fit at each $\lambda$ gives both thresholds: its late-time approach rate $\gamma_\mathrm{a}$ sets $\xi$ and hence $\obslambda{2}$ (Figs.~\fullref{fig:second_transition}{b} and~\ref{fig:second_transition_xi_fit_full}), and its first closest approach to the centre sets $R_{\min}$ and hence $\obslambda{4}$ (Fig.~\ref{fig:fr_extraction}).
    Rose bars: angular dwell-time density $\tau_\theta$, on one radial and colour scale for all panels, so that bar heights can be compared across $\lambda$.
    At $\lambda\leq0.36$ the path winds round the centre and $\tau_\theta$ is spread over all angles.
    From $\lambda=0.53$, above $\obslambdaeq{4}$, the fitted path no longer reaches $\theta=-\pi$ and a forbidden region opens, widening as $\lambda$ increases.
    Shading: forbidden region, from $\theta=-\pi$ to the most negative angle reached by the fitted path (dashed line).
    With increasing $\lambda$, $\tau_\theta$ accumulates at the angle of the fixed point, and from $\obslambdaeq{2}$ on, a single peak there dominates: the angular signature of state freezing.
    Near the edge of the forbidden region, where the path turns ($d\theta/dl=0$), $\tau_\theta$ is also enhanced by the angular parametrisation alone (Supplementary Section~\ref{App:tau_subtleties}). For this reason the transition is located from $\xi$, defined along the arc length, rather than from $\tau_\theta$.
    }
    \label{fig:second_transition_bloch_grid}
\end{figure}

The zero crossing of $\xi(\lambda)$ is located with a quadratic fit to the ten measured values nearest the sign change (Fig.~\ref{fig:second_transition_xi_fit_full}), which yields $\obslambdafull{2}$.
Uncertainties are obtained with a bootstrap: for each of $500$ replicates, we redraw the measured counts at their shot numbers and repeat the entire analysis through to the location of the crossing.
The quoted uncertainty combines the resulting $68\%$ interval with systematic contributions from the fit.
The realistic-model curve in Fig.~\fullref{fig:second_transition}{b} is shown for comparison only.
Using equation~\eqref{eq:ksi_from_eigenvalues}, $\xi$ can also be obtained from the poles of $P^{(0)}(t)$ alone (Fig.~\ref{fig:crit_lambda_1_with_poles}), giving a consistent crossing at $\lambda=0.90$.

\subsection{Opening of the Forbidden Region --- $\obslambda{4}$}
\label{App:extracting_fr_lambda}
The forbidden region opens when the no-click trajectory passes exactly through the centre of the disc (Supplementary Section~\ref{App:first_transition_splitting}).
We locate this point from the fitted trajectory $\mathbf{R}(t)$ of equation~\eqref{eq:weighted_fit}.
Let $t_{\min}$ denote the time of the first local minimum of the Bloch-vector magnitude $\|\mathbf{R}(t)\|$. We define the corresponding signed distance from the Bloch-sphere centre as
\begin{equation}
\label{eq:signed_min_distance}
R_{\min}=
-\operatorname{sgn}\!\left[\mathbf{R}(t_{\min})\cdot\mathbf{R}_\infty\right]
\|\mathbf{R}(t_{\min})\|.
\end{equation}
It is positive when the first closest point lies on the far side of the centre from the fixed point, so that the path goes round the centre, and negative when it lies on the same side, so that a wedge of angles is never visited.
Since the trajectory spirals about $\mathbf{R}_\infty$ with amplitude $A(t)=\|\mathbf{R}(t)-\mathbf{R}_\infty\|$, the closest approach occurs roughly where $\mathbf{R}(t)$ lies on the line through $\mathbf{R}_\infty$ and the centre, giving $R_{\min}\simeq A(t_{\min})-\|\mathbf{R}_\infty\|$.
Hence $R_{\min}=0$ when the amplitude left at the first pass equals the distance of the fixed point from the centre, the condition underlying equation~\eqref{eq:fr_winding_estimate}.

\begin{figure}
    \centering
    \includegraphics[width=\textwidth]{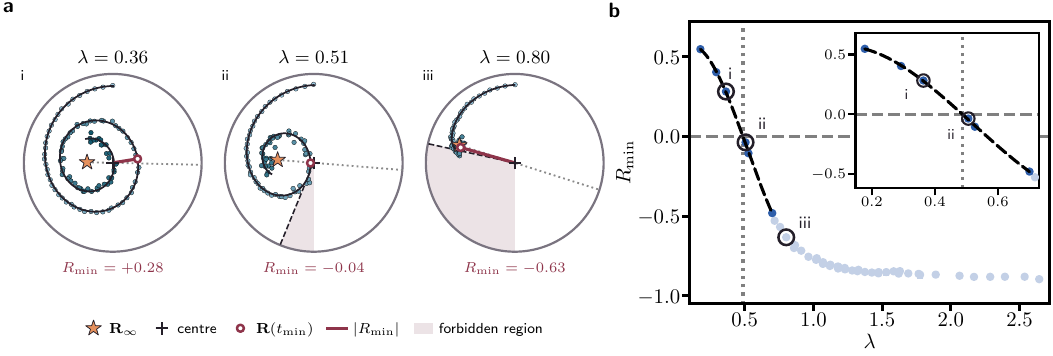}
    \caption{
    \textbf{Extraction of $\obslambda{4}$ from the signed minimum distance $R_{\min}$.}
    \textbf{a,} No-click trajectories and survival-weighted fits, as in Fig.~\ref{fig:second_transition_bloch_grid}, at three values of $\lambda$, marked i--iii in \textbf{b}.
    The maroon segment joins the centre of the disc ($+$) to the first closest approach $\mathbf{R}(t_{\min})$ ($\circ$), so that its length is $|R_{\min}|$; the star ($\bigstar$) marks the fixed point $\mathbf{R}_\infty$, and the dotted line runs from $\mathbf{R}_\infty$ through the centre.
    At $\lambda=0.36$ the closest approach lies beyond the centre, the path goes round it, and $R_{\min}>0$.
    At $\lambda=0.51$, just above $\obslambda{4}$, it lies just short of the centre, $R_{\min}=-0.04$.
    At $\lambda=0.80$ the path turns well before the centre, $R_{\min}=-0.63$, and a wedge of angles is never visited.
    Shading: forbidden region, as in Fig.~\ref{fig:second_transition_bloch_grid}; it appears at $\lambda=0.51$, where $R_{\min}$ has just turned negative, and is absent at $\lambda=0.36$.
    \textbf{b,} Signed distance $R_{\min}$ of the fitted no-click trajectory from the centre of the disc at its first closest approach, at all measured $\lambda$, with the six points nearest the sign change in full colour and the others faded; the inset shows those six points.
    Error bars are $68\%$ bootstrap intervals, smaller than the markers.
    The dashed black line shows the local cubic fit to the six points. Black circles mark the three values of $\lambda$ shown in \textbf{a} (i--iii).
    The opening of the forbidden region is identified at $\obslambdafull{4}$ (dotted grey vertical line), where the fit crosses zero (horizontal grey dashed line).
    }
    \label{fig:fr_extraction}
\end{figure}

The zero crossing of $R_{\min}(\lambda)$ is located with a cubic fit to the six measured values nearest the sign change (Fig.~\ref{fig:fr_extraction}), which yields $\obslambdafull{4}$.
Uncertainties are obtained with the same bootstrap as for $\obslambda{2}$ (Supplementary Section~\ref{App:extracting_second_transition_lambda}).
The quoted uncertainty combines the resulting $68\%$ interval with systematic contributions from the fit.

\subsection{Quantum Zeno Transition --- $\obslambda{3}$}
\label{App:extracting_QZ_transition_lambda}

\begin{figure}[h!]
    \centering
    \includegraphics{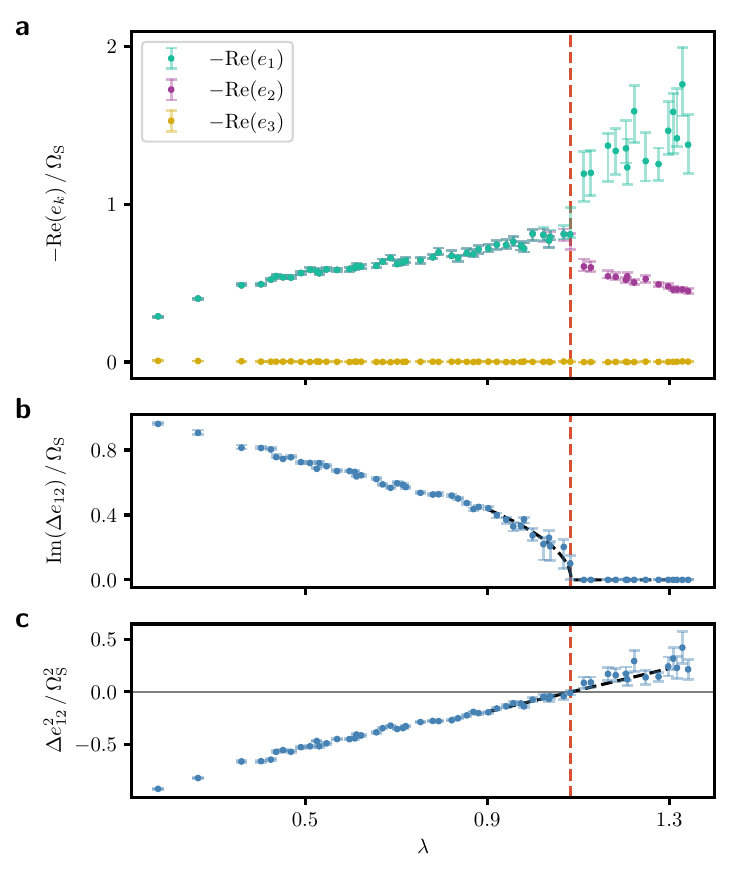}
    \caption{
    \textbf{Extracting $\obslambda{3}$ from the pole spectrum of the ensemble-averaged excited state probability $P_{\ket{1}}^\mathrm{ens}$.}
    \textbf{a,} Decay rates, given by the negative real part of the three poles governing the dynamics of $P_{\ket{1}}^\mathrm{ens}$.
    \textbf{b,} Imaginary components of the pole splitting $\Delta e_{12} = (e_{2} - e_{1})/2$ corresponding to the oscillation frequency observed in $P_{\ket{1}}^\mathrm{ens}$.
    \textbf{c,} Squared pole splitting $\Delta e_{12}^{2}$.
    The dashed black line shows a weighted linear fit of $\Delta e_{12}^{2}$ near the critical point. The dashed black curve in \textbf{b} shows $\operatorname{Im}(\Delta e_{12})$ obtained from the same fit.
    In all panels, points show estimates obtained from the original data using the matrix pencil method. Error bars are based on $68\%$ bootstrap confidence intervals, evaluated separately for each decay rate in \textbf{a}, propagated from the bootstrap distribution of $\Delta e_{12}^2$ in \textbf{b}, and symmetrised in \textbf{c}. The same bootstrap samples are used for all three panels.
    The transition is characterised by a sign change of $\Delta e_{12}^2$.
    The vertical red line marks the quantum Zeno transition at $\obslambdafull{3}$, determined from the zero crossing of the linear fit in \textbf{c}.
    }
    \label{fig:crit_lambda_3}
\end{figure}

The quantum Zeno transition is extracted from the spectral analysis of the ensemble-averaged excited state probability $P_{\ket{1}}^\mathrm{ens}$, modelled by the Liouvillian in equation~\eqref{eq:averaged-evolution-superoperator}. Although the full spectrum contains four eigenvalues, for our experimental parameters one real eigenvalue is large and negative. The corresponding eigenmode decays too rapidly to be resolved, given the experimental time resolution $T_\text{int}$, rendering it effectively unobservable. Furthermore, one eigenvalue is strictly zero, representing a steady-state solution. The remaining two eigenvalues form a complex-conjugate pair below the transition and split into two real eigenvalues above it.
This effective three-pole structure is confirmed by the singular value decomposition (SVD) step of the matrix pencil method~\cite{sarkar_using_1995}, which identifies three resolvable poles --- a zero pole and a complex-conjugate pair. We extract all three poles and track the two oscillatory poles, $e_1$ and $e_2$. Analogous to the analysis of $\obslambda{1}$, we identify the transition at the coalescence point of this pair, corresponding to the zero crossing of $\Delta e_{12}^2$, shown in Fig.~\fullref{fig:crit_lambda_3}{c}.
At each $\lambda$, two thousand bootstrap samples are generated by multinomial resampling of the possible tomography outcomes, using the experimental shot counts. The matrix pencil method is reapplied to each sample to obtain the bootstrap distributions of the poles and their splitting. The $68\%$ confidence intervals of $\Delta e_{12}^2$ are symmetrised around the original-data estimates and used to weight a linear fit following equation~\eqref{eq:approx_transition_behaviour}, with $\obslambda{1}$ replaced by $\obslambda{3}$. The fitted zero crossing yields $\obslambdafull{3}$, with the quoted uncertainty obtained from the fit.

\section{Numerical Simulations and Error Budget}
\label{App:numerical_sim}

We validate our experimental results using Monte Carlo trajectory simulations with independently measured system parameters (see Table~\ref{Tab:HamiltonianParameters}).
Each trajectory is evolved in time steps of $dt_\mathrm{sim}=\qty{10}{\ns}$. Within each time step, the qubit undergoes deterministic evolution governed by equation~\eqref{eq:theta_of_t_eq_of_motion} followed by stochastic collapse events that include dephasing, relaxation, thermal excitation, and measurement-induced transitions from $\ket{0}$ to $\ket{\mathrm{B}}$.

\begin{figure}[h!]
    \centering
    \includegraphics[width=\linewidth]{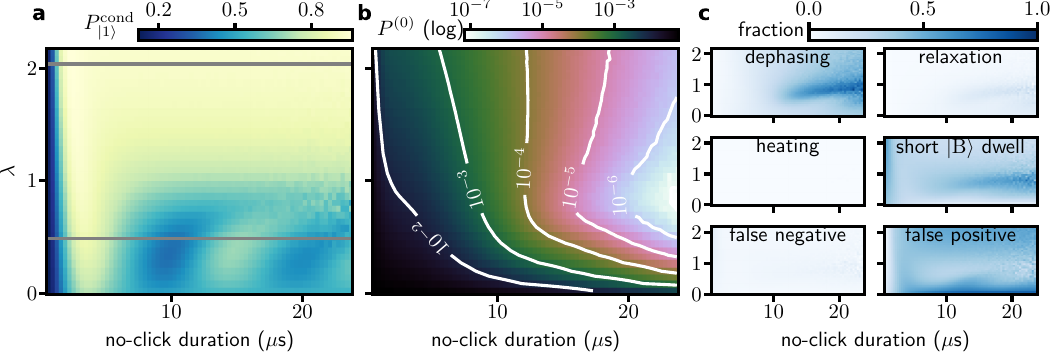}
    \caption{
    \textbf{Numerical simulation of the no-click evolution.}
    \textbf{a,} Simulated qubit excited-state population as a function of the no-click sequence duration and measurement strength~$\lambda$.
    Each pixel corresponds to the average of $3 \times 10^{8}$ stochastic trajectories, though the effective number decreases at longer durations where a click typically occurs during a trajectory (see~\textbf{b}).
    The simulation parameters are taken from Table~\ref{Tab:HamiltonianParameters} and Section~\ref{App:detector_inefficiency_realistic_model}, with an integration time of $T_\mathrm{int}=\qty{320}{\nano\second}$ to match the experimental conditions of Fig.~\fullref{fig:first_transition}{a}.
    The simulated behaviour is in good agreement with the experimental data.
    The horizontal grey lines indicate the $\lambda$-values of the one-dimensional cuts in Fig.~\fullref{fig:first_transition}{b} --- one below the quantum-jump transition ($\lambda=0.5$) and one above it ($\lambda=2.1$).
    \textbf{b,} Probability of obtaining no clicks up to time $t$, $P^{(0)}(\lambda, t)$. Most trajectories experience a click before reaching the target duration.
    \textbf{c,} Error budget. Each panel shows the fraction of trajectories that reached the target duration but experienced an event of the type indicated. 
    }
    \label{fig:numerical_simulation_and_budget_error}
\end{figure}

To emulate the experimental detection process, we group the simulation into intervals of duration ${T_\mathrm{int}}$ (corresponding to $\frac{T_\mathrm{int}}{dt_\mathrm{sim}}$ time steps) and compute the total time $t_\mathrm{B}$ spent in $\ket{\mathrm{B}}$ during each interval.
If $t_\mathrm{B} \geq \Tint/2$, the interval is  assigned to $\mathrm{B}$; otherwise, to $\bar{\mathrm{B}}$. Readout misassignments are then applied by flipping the outcome with probability $p(\bar{\mathrm{B}}|\mathrm{B})$ ($p(\mathrm{B}|\bar{\mathrm{B}})$), as obtained from the HMM analysis described in Section~\ref{App:click_statistics}. Note that a brief excursion to $\ket{\mathrm{B}}$ lasting less than $\Tint /2$ results in an undetected excursion before the readout misassignment is applied, accounting for the mechanism discussed in Supplementary Section~\fullref{App:realistic_model}{C}.
Fig.~\ref{fig:numerical_simulation_and_budget_error} presents simulation results corresponding to the experimental data in Fig.~\ref{fig:first_transition}.
Panel~\textbf{a} shows the simulated qubit excited-state population $P^{\mathrm{cond}}_{\ket{1}}$ conditioned on no clicks.
Panel~\textbf{b} displays the no-click survival probability $P^{(0)}(\lambda, t)$ and panel~\textbf{c} summarises the error budget by showing the fraction of trajectories affected by each error type.

\clearpage
\twocolumngrid
\end{document}

%% file: lambda_macros.tex
\newcommand{\translabel}[1]{%
    \ifcase#1\relax
        #1%
    \or \mathrm{QJ}%
    \or \mathrm{SF}%
    \or \mathrm{QZ}%
    \or \mathrm{FR}%
    \else #1%
    \fi
}

\newcommand{\baselam}[1]{\lambda}

\newcommand{\generallambda}[1]{\baselam{#1}_{\translabel{#1}}}
\newcommand{\critlambda}[1]{\baselam{#1}^\mathrm{ideal}_{\translabel{#1}}}
\newcommand{\reallambda}[1]{\baselam{#1}^\mathrm{realistic}_{\translabel{#1}}}
\newcommand{\obslambda}[1]{\baselam{#1}^\mathrm{exp}_{\translabel{#1}}}
\newcommand{\obslambdaval}[1]{%
    \ifcase#1\relax #1%
        \or 1.30%
        \or 0.90%
        \or 1.09%
        \or 0.49%
    \else ??\fi}
\newcommand{\obslambdaerr}[1]{%
    \ifcase#1\relax
        \or \pm0.04%
        \or \pm0.01%
        \or \pm0.01%
        \or \pm0.01%
    \else {}\fi}
\newcommand{\reallambdaval}[1]{%
    \ifcase#1\relax #1
    \or 1.27
    \or 1.00
    \or 1.20
    \or 0.52
    \else ??\fi
    }

\newcommand{\obslambdaeq}[1]{\obslambda{#1}=\obslambdaval{#1}}

\newcommand{\obslambdafull}[1]{\obslambdaeq{#1}\obslambdaerr{#1}}